%% file: main.tex
\documentclass{article}
\usepackage{xcolor}
\usepackage{graphicx} % Required for inserting images
\usepackage{amsmath}
\usepackage{amssymb}
\usepackage{bbm}
\usepackage{authblk} % For affiliations
\usepackage{tabularx} %table S1
\usepackage{makecell} %table S1
\usepackage{lineno}
\usepackage[final]{changes}
\usepackage{booktabs}
\usepackage{multirow}
\usepackage{siunitx}    % For S column type (decimal alignment)
\usepackage{ragged2e} %justify text below figures
\usepackage{subcaption}

\usepackage[style=numeric-comp, sorting=none, backend=bibtex, doi=false, eprint=false, isbn=false, url=false]{biblatex}
\AtEveryBibitem{%
  \clearfield{url}       % Removes URL
  \clearfield{urldate}   % Removes 'Accessed on'
  \clearfield{doi}       % Removes DOI
  \clearfield{eprint}    % Removes arXiv or other preprint info
  \clearfield{note}      % Removes additional notes
  \clearfield{issn}      % Removes ISSN
  \clearfield{isbn}      % Removes ISBN
  \clearfield{pagetotal} % Removes total number of pages
  \clearfield{language}  % Removes language field
  \clearlist{location}   % Removes publisher location
  \clearlist{publisher}  % Removes publisher (for articles, not needed)
}

\title{A statistical framework for comparing epidemic forests}
\author[1,2,3]{Cyril Geismar}
\author[2,3,4]{Peter J. White}
\author[2,3, *]{Anne Cori}
\author[2,3, *]{Thibaut Jombart}
\affil[1]{Bloomberg School of Public Health, Johns Hopkins University, Baltimore, United States}
\affil[2]{MRC Centre for Global Infectious Disease Analysis, Imperial College School of Public Health, London, United Kingdom}
\affil[3]{NIHR Health Protection Research Unit in Modelling and Health Economics, Imperial College School of Public Health, London, United Kingdom}
\affil[4]{UK Health Security Agency, London, United Kingdom}
\affil[*]{authors contributed equally}
\date{} 

\begin{document}

\maketitle
\section{Abstract}
Inferring who infected whom in an outbreak is essential for characterising transmission dynamics and guiding public health interventions. However, this task is challenging due to limited surveillance data and the complexity of immunological and social interactions. Instead of a single definitive transmission tree, epidemiologists often consider multiple plausible trees forming \textit{epidemic forests}. Various inference methods and assumptions can yield different epidemic forests, yet no formal test exists to assess whether these differences are statistically significant. We propose such a framework using a chi-square test and permutational multivariate analysis of variance (PERMANOVA). We assessed each method\replaced{'s ability to distinguish simulated epidemic forests generated under different offspring distributions. While both methods achieved perfect specificity for forests with 100+ trees, PERMANOVA consistently outperformed the chi-square test in sensitivity across all epidemic and forest sizes.}{on simulated epidemic forests with and without superspreading, assessing their ability to detect differences between two epidemic forests. Both methods demonstrated near-perfect sensitivity, but PERMANOVA outperformed the chi-square test with near-perfect specificity for forests with 100+ trees.} Implemented in the R package \textit{mixtree}, we provide the first statistical framework to robustly compare epidemic forests.
\clearpage
\section{Author Summary}
\added{Identifying who infected whom is a central part of outbreak investigation. It helps trace the source of infection, uncover missing cases, identify superspreaders, and describe broader dynamics of transmission such as its speed, pattern, and scale. With the advent of pathogen sequencing and digital contact tracing, computational models have become the standard approach for reconstructing outbreaks. These probabilistic models do not identify a single definitive history of who infected whom (\textit{i.e.} a transmission tree), but a collection of plausible alternatives, which we call `epidemic forests'. Different modeling assumptions or data sources can produce different epidemic forests, but until now, there has been no formal way to determine whether these differences are meaningful.\\
We present the first statistical framework designed to compare epidemic forests. We evaluate two methods: one that counts how often specific transmission pairs appear, and another that compares the structure of transmission trees. Testing these methods on simulated outbreaks, we found that both successfully identified when forests represented identical transmission dynamics, but one method outperformed the other in identifying forests representing distinct transmission dynamics. Our framework, implemented in the R package mixtree, enables epidemiologists to validate and compare outbreak reconstruction approaches, supporting more reliable investigations.}
\clearpage
\section{Introduction}
Tracking who infected whom is central to outbreak investigations. Transmission trees, modelled as directed acyclic graphs (DAGs) where vertices represent infected individuals and directed edges indicate transmission events, delineate infector-infectee relationships \cite{jombart_reconstructing_2011}. These representations can assist epidemiologists in identifying introduction and superspreading events \cite{wang_inference_2020,frieden_identifying_2020}, whilst also elucidating broader transmission dynamics relevant to outbreak response. The topology of transmission trees, defined by the arrangement of vertices and edges, encodes key epidemiological parameters.  The out-degree distribution of vertices represents the number of secondary infections per infected individual (\textit{i.e. }the offspring distribution), revealing the degree of heterogeneity in transmission \cite{lloyd-smith_superspreading_2005, abbas_reconstruction_2022, didelot_bayesian_2014, didelot_genomic_2017}. Branching patterns inform on transmission dynamics between groups \cite{geismar_sorting_2024}, revealing group reproduction numbers \cite{abbas_reconstruction_2022} and transmission patterns, for example, between healthcare workers and patients in nosocomial outbreaks \cite{abbas_explosive_2021} or between children and adults in schools \cite{kremer_reconstruction_2023}. \hfill \break

The inference of transmission trees is challenging and often characterised by large uncertainty, partly due to \added{the lack of discriminatory power in choosing between possible transmission pairs \cite{campbell_when_2018, abbas_explosive_2021, abbas_reconstruction_2022}}, incomplete surveillance data and diverse, sometimes conflicting sources (\textit{e.g.} contact, temporal, spatial, or genetic data) \cite{duault_methods_2022}. Additional complexities arise from varying methodological approaches \cite{duault_methods_2022} and pathogen evolution mechanisms that are difficult to model (\textit{e.g.} within-host evolution and transmission bottleneck \cite{didelot_genomic_2017, de_maio_scotti_2016}). Consequently, outbreak reconstruction often yields \textit{epidemic forests}, which are collections of plausible transmission trees rather than a singular definitive representation of who infected whom. \hfill \break

Without formal statistical methods to differentiate epidemic forests, determining whether differences between them represent meaningful variations in transmission dynamics or \deleted{are due to sampling and model} uncertainty \added{in tree reconstruction} is challenging. Such distinction would help validate convergence when repeated model runs produce statistically similar forests and assess whether competing inference approaches or alternative data sources yield significantly different forests.\hfill \break

Bayesian inference methods have emerged as the gold standard for transmission tree reconstruction, with various approaches differing in their assumptions, data requirements, and inference strategies \cite{duault_methods_2022}. In this context, epidemic forests represent samples drawn from a model's posterior distribution of transmission trees.  To assess \replaced{the performance of the inference process}{model reliability and compare competing approaches}, researchers rely on general Markov Chain Monte Carlo (MCMC) diagnostics, applied to scalar parameter chains rather than the inferred trees themselves. These diagnostics evaluate convergence through trace plot inspection and the Gelman-Rubin statistic \cite{gelman_inference_1992},  assess sampling efficiency through effective sample size calculations, and check model fit using posterior predictive checks \cite{lambert_students_2018}. In parallel, \textit{consensus trees} are used to summarise epidemic forests, typically representing \replaced{, for each case, the infector with the highest posterior support across samples}{the most common infector-infectee pairs across samples }\cite{felsenstein_confidence_1985, jombart_treespace_2017, jombart_bayesian_2014, klinkenberg_simultaneous_2017, campbell_when_2018}.  However, these trees are often abstract representations rather than plausible transmission scenarios, potentially introducing cycles or multiple index cases\added{.While algorithms such as Edmonds can enforce a valid tree topology, the resulting consensus tree may correspond to a combination of ancestries that was never observed as a complete tree in the posterior} \cite{klinkenberg_simultaneous_2017, gibbons_algorithmic_1985, hall_epidemic_2015}. \hfill\break

Consequently, standard MCMC diagnostics assess parameter chains rather than the inferred transmission events, while consensus trees ignore the uncertainty in who infected whom and may misrepresent key epidemiological features. This underscores the need for specialised statistical methods that can differentiate epidemic forests while accounting for uncertainty and relevant topological properties.\hfill\break

\begin{figure}[!h]
    \centering
    \includegraphics[width=1\linewidth]{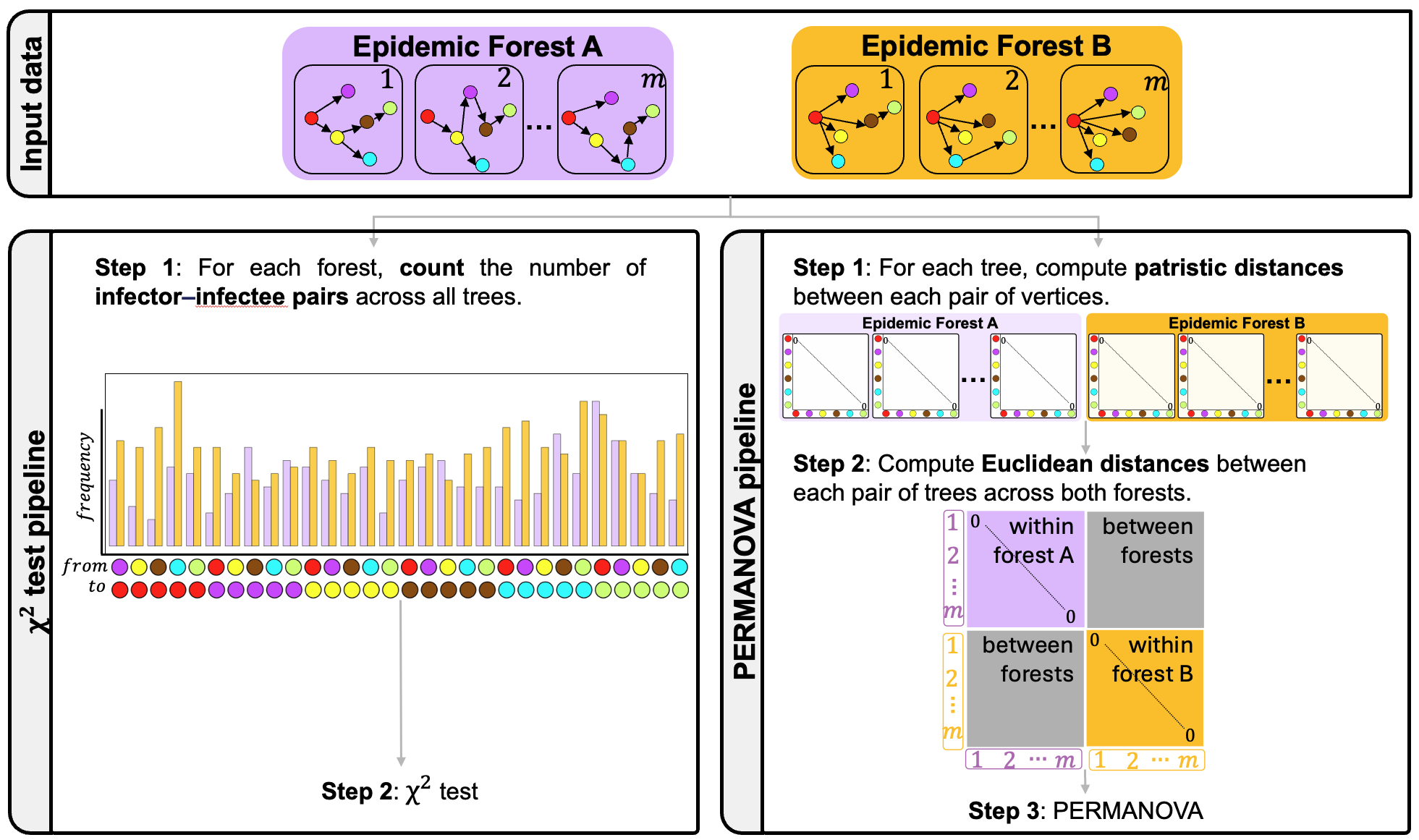}
    \caption{Statistical framework for comparing epidemic forests.}
    \medskip
    \small
    \justifying
    Diagram illustrating the methods for comparing epidemic forests $A$ (pink, \added{\textit{e.g.}} no superspreading) and $B$ (orange, \added{\textit{e.g.}} superspreading). Coloured dots represent infected cases, and arrows indicate transmission events. \linebreak
    Left: The $\chi^2$ test compares the frequency of infector-infectee pairs between forests. \linebreak
    Right: The PERMANOVA method first calculates pairwise \replaced{graph}{patristic} distances (number of transmission events between two cases, plus one - see Fig. \ref{fig:patristic}) between vertices within each tree, converts these to a Euclidean distance matrix between all trees, and tests for significant topological differences between the forests using permutation-based testing.\linebreak 
    Both methods test the null hypothesis that the compared epidemic forests stem from the same generative process.
    \label{fig:diagram}
\end{figure}

Here, we introduce a statistical framework for testing differences between epidemic forests. We consider two alternative methods: a chi-square ($\chi^2$) test \cite{pearson_x_1900} used to compare the frequency of  infector-infectee pairs between forests, and a permutation-based multivariate analysis of variance (PERMANOVA) \cite{anderson_new_2001} which compares topological distances between trees within and between forests. Both methods are summarised in Figure \ref{fig:diagram}. We evaluated the performance of each method by comparing simulated epidemic forests with \replaced{varying offspring distributions}{and without superspreading events}, measuring their ability to correctly identify forests stemming from distinct (sensitivity) or identical (specificity) generative processes (see Methods).\hfill\break

\clearpage
\section{Results}
We simulated pairs of epidemic forests that were either stemming from the same, or different generative processes (see Methods). We systematically varied epidemic size ($\epsilon$: 20–200 cases), forest size ($m$: 20–200 trees per \replaced{forest}{sample}), and the \replaced{parameters of the negative binomial offspring distribution ($R_0$: 1.5-3 and $k$: 0.1-Poisson-like), which determine the mean and dispersion of secondary infections.}{\textit{overlap frequency} ($\omega$: 0–1), a parameter gradually controlling the similarity between the processes generating the two forests with $\omega=0$ for distinct processes and $\omega=1$ for identical processes.} \deleted{For each of the 96 parameter combinations,} We simulated \replaced{90,000}{10,000 pairs of} epidemic forests (see Methods) based on which \added{we conducted 5,760,000 tests to measure} sensitivity and specificity  for the $\chi^2$ test and PERMANOVA. \hfill \break

\begin{figure}[!ht]
    \centering
    \includegraphics[width=1\linewidth]{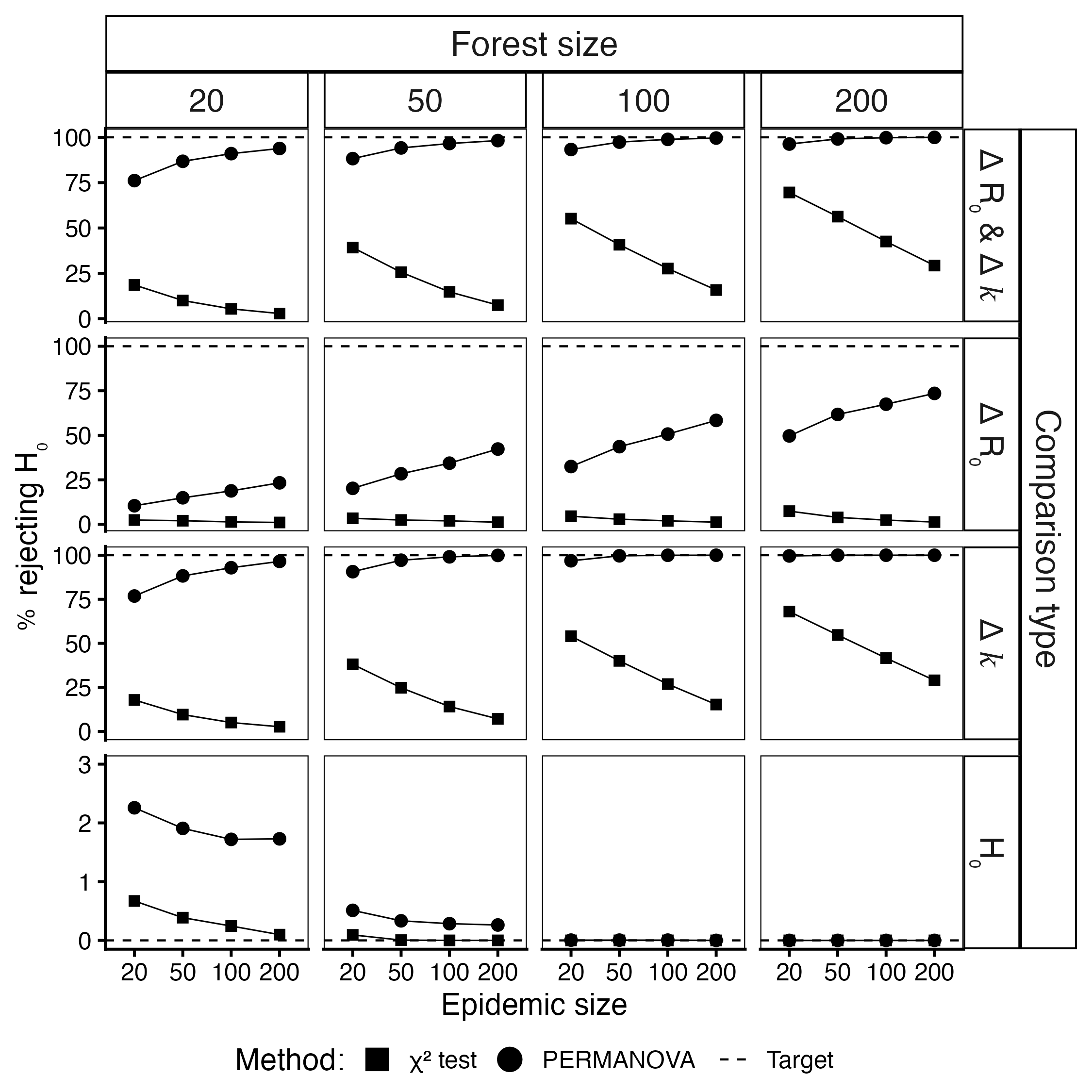}
    \caption{Performance of $\chi^2$ and PERMANOVA for comparing epidemic forests.}
    \medskip
    \small
    \justifying
     This figure summarises simulation results for the $\chi^2$ test \replaced{(square points)}{(green)} and PERMANOVA \replaced{(circular points)}{(red)} across \replaced{varied parameter conditions}{increasing levels of similarity between epidemic forests}. The y-axis shows the percentage of tests rejecting the null hypothesis \added{($H_0$)} of no difference between forests. The x-axis displays the \added{epidemic sizes. Column panels refer to the forest size (\textit{i.e.} the number of trees in each forest). Row panels refer to the type of differences between the two forests (with $H_0$ for no differences).} \deleted{overlap frequency ($\omega$), which controls the proportion of trees drawn from the same generative process between the two epidemic forests being compared.
     Black asterisks mark perfect results: 100\% rejection for $\omega < 1$, and 5\% rejection for $\omega = 1$ (at $\alpha = 0.05$). Panels are arranged by epidemic size (columns: 20–200 cases) and forest size (rows: 20–200 trees).}
    \label{fig:power_curves.png}
\end{figure}

\added{Overall results are presented as Receiver-Operator Characteristic (ROC) curves (supplementary Fig. \ref{fig:roc-curves})}, with area under the curve (AUC) provided in supplementary Figure \ref{fig:auc}.  Both methods exhibited near perfect \replaced{specificity ($>$ 97\%)}{sensitivity ($>$ 95\%)} \deleted{in the presence of substantial differences between the simulated epidemic forests ($\omega<$ 0.6)}, \added{ \textit{i.e.} the ability to correctly identify forests drawn from identical generative processes,} across all epidemic or forest sizes (Fig.\ref{fig:power_curves.png}, \added{row 4}).\deleted{ As differences between simulated forests decreased  (0.6 $\leq \omega \leq $ 0.8), the $\chi^2$ test displayed better sensitivity than PERMANOVA for small forest sizes ($m$ = 20), while both methods maintained similarly high sensitivity with larger forest sizes ($m$ = 200) (Fig.\ref{fig:target_percentage.png})}. \added{The $\chi^2$ test had a negligible advantage in specificity (+1.5\%) when the number of trees in each forest was small ($m\leq 50$). 
Across the aggregated simulation results, sensitivity was near perfect once the forest size reached 50-100 trees, with AUC nearing 1 (supplementary Fig. \ref{fig:roc-curves}). However, these results varied substantially between methods and simulation settings.} \hfill \break

The methods differed substantially in their \replaced{sensitivity}{specificity}, \textit{i.e.} their ability to correctly identify forests drawn from \replaced{different}{identical} generative processes\deleted{($\omega$=1)}\added{, with PERMANOVA consistently outperforming the $\chi^2$ test across all scenarios (Fig.\ref{fig:power_curves.png}, row 1-3)}. \deleted{PERMANOVA maintained near-perfect specificity across all $\alpha$ thresholds, irrespective of epidemic or forest size (Fig.\ref{fig:target_percentage.png}, supplementary information Fig. \ref{fig:roc-curves}). In contrast, the $\chi^2$ test exhibited decreasing specificity as epidemic size increased, with false positive rates escalating up to 23\% for large outbreaks ($\epsilon$ = 200) at the conventional $\alpha$ = 0.05 level (Fig.\ref{fig:target_percentage.png}). This poor specificity in the $\chi^2$ test persisted across all significance thresholds (supplementary information Fig. \ref{fig:roc-curves}), with the inverse relationship between epidemic size and specificity undermining its reliability for moderate to large outbreaks ($\epsilon \geq$ 50) (Fig.\ref{fig:target_percentage.png}, Fig.\ref{fig:auc}).} \added{However, the magnitude of PERMANOVA's advantage varied considerably depending on which parameters differed between forests. A logistic regression model explained 58\% of the variance in test sensitivity (Pseudo $R^2$ \cite{tjur_coefficients_2009} = 0.58), with method choice, forest size, epidemic size, and differences in $R_0$ and $k$ as key predictors (Table \ref{tab:glm-sensitivity}). Compared to the $\chi^2$ test, PERMANOVA showed much greater sensitivity when forests differed in their dispersion parameter ($\Delta_{k}$),  with 51-fold higher odds of correctly distinguishing overdispersed from Poisson-like forests ($\Delta_{k_{(0, 1] \text{ vs. Poisson}}}$), and 8-fold higher odds when comparing forests with different degrees of overdispersion ($\Delta_{k_{(0, 1] \text{ vs.}(0, 1]}}$) (Table \ref{tab:glm-sensitivity}). When forests differed in dispersion and contained at least 100 trees, PERMANOVA achieved near-perfect sensitivity (98.7\%), irrespective of epidemic size (Fig. \ref{fig:power_curves.png}, rows 1 and 3, column 3).\\
In contrast, PERMANOVA's advantage over the $\chi^2$ test diminished when forests differed only in reproduction number (OR = 3, Table \ref{tab:glm-sensitivity}). When both forests shared strong overdispersion (common $k \leq 0.5$), high stochastic variability in individual transmission limited the ability to detect differences in $R_0$ up to 1 ($\Delta R_0 \leq 1$), yielding low sensitivity even with 200 trees per forest (52\% across epidemic sizes; supplementary Figures  \ref{fig:grid_combined} and \ref{fig:delta_R0_lineplots_m200}). Sensitivity improved progressively as the common dispersion parameter approached Poisson-like transmission ($k \to \infty$) or as epidemic size increased (supplementary Fig. \ref{fig:delta_R0_lineplots_m200}).\\
In addition to higher sensitivity, PERMANOVA produced consistently narrower p-value distributions than the $\chi^2$ test, with interquartile ranges substantially smaller across all scenarios (supplementary Fig. \ref{fig:delta_pval.png}).\\
Both methods’ sensitivity increased with forest size (OR = 3, 6, 12 for $m$= 50, 100 and 200 respectively) but showed opposite correlations with epidemic size: PERMANOVA’s sensitivity rose with larger epidemics (OR=  2, 4 and 5  for $\epsilon$= 50, 100, and 200 respectively), whereas the $\chi^2$ test's sensitivity declined (OR= 0.5, 0.3, and 0.1) (Table \ref{tab:glm-sensitivity}, Fig. \ref{fig:power_curves.png}).}\hfill \break
Our findings establish PERMANOVA as the \replaced{superior}{preferred} method for comparing epidemic forests when sufficient samples are available ($m \geq$ 100), providing excellent sensitivity and specificity regardless of epidemic size. \deleted{(AUC $>$ 0.999, Fig. \ref{fig:auc}). The $\chi^2$ test remains valuable primarily for scenarios involving small epidemics ($\epsilon < 50$) with small forests ($m \leq 20$) (Fig. \ref{fig:target_percentage.png}, Fig. \ref{fig:auc}).}

% -------------------------
\begin{table}[htbp]
\centering
\caption{Logistic regression results for the sensitivity model (Eq.\ref{eq:sensitivity})}
\label{tab:glm-sensitivity}
\begin{tabular*}{\linewidth}{@{\extracolsep{\fill}} l l S[table-format=3.2] S[table-format=<1.3] }
\toprule
\multicolumn{2}{l}{\textbf{Predictor}} & {\text{\textbf{Odds Ratio}}} & {\textbf{$p\text{-value}$}} \\
\midrule
Intercept & & 0.01 & {$<$0.001} \\
\addlinespace
\textbf{Method} & PERMANOVA & 1.94 & {$<$0.001} \\
\addlinespace
\multirow{3}{*}{\textbf{Forest size }($m$)} 
 & 50 & 2.93 & {$<$0.001} \\
 & 100 & 6.10 & {$<$0.001} \\
 & 200 & 12.06 & {$<$0.001} \\
\addlinespace
\multirow{3}{*}{\textbf{Epidemic size} ($\varepsilon$)} 
 & 50 & 0.54 & {$<$0.001} \\
 & 100 & 0.30 & {$<$0.001} \\
 & 200 & 0.15 & {$<$0.001} \\
\addlinespace
\multirow{3}{*}{\textbf{Parameter difference} ($\Delta$)}\\
\addlinespace
 & $R_0$ & 1.70 & {$<$0.001} \\
& $k$ & & \\
& \quad $(0, 1]$ vs. $(0, 1]$ & 33.04 & {$<$0.001} \\
& \quad $(0, 1]$ vs. Poisson & 36.74 & {$<$0.001} \\
& $R_0 : k$ & & \\
& \quad\quad $(0, 1]$ vs. $(0, 1]$ & 0.62 & {$<$0.001}\\ 
& \quad\quad $(0, 1]$ vs. Poisson & 0.61 & {$<$0.001}\\ 
\multirow{3}{*}{\textbf{PERMANOVA : $\varepsilon$}}\\
 & 50 & 4.19 & {$<$0.001} \\
 & 100 & 11.85 & {$<$0.001} \\
 & 200 & 33.94 & {$<$0.001} \\
\addlinespace
\multirow{3}{*}{\textbf{PERMANOVA : $\Delta$}}\\
\addlinespace
 & $R_0$ & 3.01 & {$<$0.001} \\
 & $k$ & & \\
& \quad $(0, 1]$ vs. $(0, 1]$ & 7.77 & {$<$0.001} \\
& \quad $(0, 1]$ vs. Poisson & 51.38 & {$<$0.001} \\
& $R_0 : k$ & & \\
& \quad \quad $(0, 1]$ vs. $(0, 1]$ & 0.21 & {$<$0.001}\\ 
& \quad \quad $(0, 1]$ vs. Poisson & 0.15 & {$<$0.001}\\ 
\addlinespace
\bottomrule
\multicolumn{4}{p{0.9\linewidth}}{\footnotesize \textit{Note:} Pseudo $R^2$ = 0.58 \cite{tjur_coefficients_2009}. All p-values are $<0.001$ due to the large simulation sample size ($n =5,040,000$). `:' denotes the interaction term. The reference categories are: $\chi^2$test for method, 20 for $\varepsilon$ and $m$, 0 for $\Delta k$.}
\end{tabular*}
\end{table}
\clearpage
\section{Discussion}
We evaluated two statistical approaches, \added{ the chi-square ($\chi^2$) test and PERMANOVA,}  for distinguishing between collections of transmission trees (\textit{i.e.} epidemic forests) originating from different generative processes\deleted{: the chi-square ($\chi^2$) test and PERMANOVA} \added{, defined by the mean and dispersion of their offspring distribution (\textit{i.e.} the distribution of secondary cases generated by each infected individual)}.  The \replaced{$\chi^2$ test}{former} tests for differences in the frequency of infector-infectee pairs between epidemic forests, treating each pair as isolated edges without considering their relative position within each tree.  In contrast, \replaced{PERMANOVA}{the latter} leverages customised tree-based distance metrics to quantify meaningful epidemiological differences between tree topologies, which may better signal distinct pathogen transmission dynamics. 

Our simulations showed that PERMANOVA consistently outperformed the $\chi^2$ test \deleted{when comparing epidemic forests} \added{in distinguishing epidemic forests generated under different offspring distributions} \deleted{comprising 100 or more transmission trees}. It achieved near-perfect sensitivity \deleted{and specificity (AUC$>$0.99)} \added{when forests differed in their dispersion parameter} across all epidemic sizes (20-200 cases), provided forests contained at least 100 trees. \added{However, its performance declined when forests differed solely in their mean reproduction number, especially for forests with high overdispersion (common $k < 0.5$) (supplementary Fig. \ref{fig:delta_R0_lineplots_m200}). In such settings, the substantial stochastic variability in individual transmission masked differences in mean transmissibility.} Although the $\chi^2$ test also demonstrated excellent \replaced{specificity}{sensitivity}, its \replaced{sensitivity was consistently lower across all scenarios and declined further as epidemic size increased.}{specificity and further declined progressively as outbreak size increased, limiting its reliability for larger epidemics. The $\chi^2$ test remains valuable primarily for small outbreaks ($<$ 50 cases) with limited samples ($\leq$ 20 trees).} \added{Larger epidemics produced higher forest entropy, which indicates greater variation in who infected whom across trees (supplementary Fig.\ref{fig:entropy.png}, Table\ref{tab:lm-entropy}). Increased entropy yielded sparse contingency tables with many low expected counts and growing degrees of freedom, which reduced the statistical power of the $\chi^2$ test (see Methods, Eq.\ref{eq:chisq}). In contrast, PERMANOVA became more sensitive as epidemics grew given that additional transmission events reduced the variance in within group distances, increasing the F-statistic (see Methods, Eq.\ref{eq:f-statistic}).}

Computationally, both methods scale with epidemic size, although PERMANOVA incurs greater computational expense (see supplementary material). Parallelisation and constrained permutation (for PERMANOVA \cite{oksanen_vegan_2025}) or replicates used in the Monte Carlo test (for $\chi^2$ test \cite{r_core_team_r_2025}) make both methods applicable to most contexts. When comparing two forests, each with 100 trees and 100 vertices, the  $\chi^2$ test takes 0.5 seconds, while PERMANOVA takes an average of 5 seconds (supplementary material Table \ref{tab:benchmark_table}). \hfill \break  To facilitate accessibility of these methods, we have developed \textit{mixtree} \cite{geismar_mixtree_2025}, a free, open-source R package available on CRAN \cite{r_core_team_r_2025}. \textit{mixtree} implements both the $\chi^2$ test and PERMANOVA methods described in this study. \hfill \break

The proposed framework addresses several needs for outbreak reconstruction. First, it provides a formal approach for assessing MCMC convergence in `tree space' by comparing epidemic forests sampled from independent MCMC chains, which should be statistically indistinguishable when converged. This method complements existing diagnostics that focus on scalar parameter chains, which do not fully capture the complex tree structures that form the primary output of Bayesian inference models. Second, it enables rigorous comparison between competing models with different assumptions about transmission dynamics, facilitating evidence-based model selection. Third, it can detect whether incorporating additional data sources (\textit{e.g.} contact tracing \cite{campbell_bayesian_2019}) into reconstruction efforts significantly alters the resulting transmission trees, helping researchers evaluate the value of supplementary data. However, it cannot independently determine which reconstruction is more accurate without additional validation measures. \hfill \break 

Our study focused on comparing two forests of equal size for computational feasability. However, both methods can compare any number of forests of varying sizes sharing the same set of vertices, as implemented in our \textit{mixtree} package. Nonetheless, the two methods do not share identical limitations. PERMANOVA assumes full graph connectivity \cite{bang-jensen_connectivity_2009} , so it cannot accommodate multiple introductions that result in disconnected trees. In contrast, the $\chi^2$ test can handle multiple introductions by assigning them to a dedicated category in the edge list \deleted{(Fig. \ref{fig:diagram})}. In the presence of unobserved cases, the $\chi^2$ test cannot distinguish between direct and unobserved intermediate transmissions. Importantly, PERMANOVA could be extended by modelling epidemiological, spatial or genetic distances as edge weights. For example, these weights could represent the number of infection generations between pairs of cases, thus accounting for unobserved cases. The simple \replaced{graph}{patristic} distance used here could be replaced with a more complex metric that incorporates additional edge characteristics (\textit{i.e.} weights) such as the number of generations between observed cases, or the time difference between their symptoms \cite{geismar_bayesian_2023} or infection dates. \added{While our simulation framework assessed method performance when the forest's generative process differed only in its offspring distribution, other epidemic features also shape tree topology. Future work should evaluate performance under alternative assumptions about epidemic dynamics such as group transmission patterns \cite{geismar_sorting_2024}, the effects of saturation \cite{cori_new_2013}, vaccination or new variants of concern \cite{geismar_bayesian_2023}, which would require developing additional distance metrics for PERMANOVA to capture such features. Our simulation framework focused on epidemics of 20–200 cases, reflecting the typical range for computational outbreak reconstruction, and our results show that PERMANOVA performs well once forests comprise 100 or more trees, corresponding to the typical effective sample size from Bayesian reconstruction models \cite{duault_methods_2022}.}\hfill \break 

\added{While alternative methods for comparing graph collections exist, they typically rely on abstract graph kernels not directly interpretable in our epidemiological context \cite{gudmundarson_gtst_2024}. In contrast, our method employs a distance metric that is epidemiologically meaningful as it corresponds to the number of generations of infection separating each pair of cases. Furthemore, PERMANOVA can also be used for multifactorial analysis to quantify the relative contributions of the inference method, data type, and prior assumptions to the observed topological differences between epidemic forests.} In addition to the application to epidemic reconstruction that we have considered here, this work addresses a more general methodological gap across disciplines where relational structures are represented as graphs \cite{gross_graph_2018, chen_entity-relationship_1976, zerbino_velvet_2008, balaban_applications_1985, granovetter_strength_1973, drummond_beast_2007}. In practice, diverse data sources, modelling assumptions, and analytical methods typically produce not single solutions but ensembles of plausible alternatives, \textit{i.e.} collections of graphs. Bayesian approaches excel at generating these collections through MCMC sampling but lack formal statistical tools for comparing the resulting posterior samples. One example of other such application area is phylogenetic tree reconstruction \cite{drummond_beast_2007}, where researchers encounter similar challenges that can lead to conflicting evolutionary hypotheses or taxonomic classifications. In information and network science, different network representations may likewise suggest distinctive social patterns or information flow dynamics. \hfill \break

In conclusion, our framework enables the comparison of collections of transmission trees, a special class of graph, by distinguishing meaningful structural variations from sampling and model uncertainty. We have demonstrated its utility to epidemic reconstruction, but this approach likely extends to other fields relying on graph-based representations. We encourage researchers to adapt and validate this framework to address domain-specific challenges in their respective fields, potentially developing additional metrics that capture the unique characteristics of their data structures.
\clearpage
\section{Methods}
We introduce a framework for comparing collections of transmission trees, termed \textit{epidemic forests}. We present two approaches: the first based on a $\chi^2$ test \cite{pearson_x_1900} on transmission pair frequencies, and the second using PERMANOVA, a method originally developed for ecological community analysis \cite{anderson_new_2001}, on transmission tree distances. Both methods are described below and illustrated in Fig. \ref{fig:diagram}. We use a simulation to compare the respective performances of the two approaches. 

\subsection{Epidemic Forests}
Transmission trees represent the spread of a disease amongst infected individuals as directed acyclic graphs (DAGs) \cite{jombart_reconstructing_2011}.  A transmission tree $ T = (V, E) $ consists of a set $V = \{v_1, v_2, \ldots, v_n\}$ containing $n$ vertices (each representing an infected individual) and a set $E = \{e_2, e_3, \ldots, e_{n}\}$ of $n-1$ directed edges. Each edge represents an infector-infectee pair, denoted as $e_j = (v_i, v_j)$, with $v_i, v_j \in V$ and $v_i \neq v_j $. This directed edge connects an infector $v_i$ to its infectee $v_j$, formally encoding the `who infected whom' relationship. All vertices have an in-degree of 1, except the root which represents the index case and has an in-degree of 0. In the absence of data to define meaningful edge weights, we assume all edges have a weight of 1.

We define an \textit{epidemic forest} as a collection of transmission trees, each with the exact same set of vertices, but possibly different sets of edges. We consider two epidemic forests $\mathcal{F}_A = (T^A_1, \ldots, T^{A}_{m_A})$ and 
$\mathcal{F}_B = (T^B_1, \ldots, T^{B}_{m_B})$, where the $k^{\text{th}}$ tree in  $\mathcal{F}_A$ is defined as $T^A_k = (V, E^A_k)$.
For simplicity, we assume that the two epidemic forests have the same size ($m_A=m_B = m$), but the approaches described below can readily accommodate ($m_A \neq m_B$).\added{ In practice, an epidemic forest may be obtained by sampling from a posterior distribution via Bayesian inference (\textit{e.g.}, MCMC) or from a stochastic transmission model \cite{duault_methods_2022, watson_probability_1875}.}

\subsection{$\chi^2$ test}
The $\chi^2$ test compares the absolute frequencies of infector-infectee pairs (\textit{i.e.} edges) between two epidemic forests $\mathcal{F}_A$ and $\mathcal{F}_B$. For each of the possible infector-infectee pair, we count their occurrences across all trees in a forest $\mathcal{F}_X$ as:
\begin{equation}
c^{\mathcal{F}_X}_{ij} = \sum_{l=1}^{m} \mathbbm{1}_{((v_i, v_j) \in E^X_l)}
\end{equation}
where $\mathbbm{1}$ is the indicator function (yielding 1 if the pair appears in tree $T^X_l$, 0 otherwise).

The $\chi^2$ statistic for comparing forests $\mathcal{F}_A$ and $\mathcal{F}_B$ is:
\begin{equation}
\chi^2 = \sum_{(i,j) \in \mathcal{P}} \frac{(c^{\mathcal{F}_A}_{ij} - c^{\mathcal{F}_B}_{ij})^2}{c^{\mathcal{F}_A}_{ij} + c^{\mathcal{F}_B}_{ij}}
\label{eq:chisq}
\end{equation}
where $\mathcal{P} = \{(i, j) \mid i \neq j, c^{\mathcal{F}_A}_{ij} + c^{\mathcal{F}_B}_{ij} > 0\}$ includes only infector-infectee pairs observed in at least one forest. Under the null hypothesis that both forests stem from the same underlying frequency distribution of infector-infectee pairs, $\chi^2$ follows a chi-square distribution with $|\mathcal{P}| - 1$ degrees of freedom, where $|\mathcal{P}|$ denotes the number of unique infector-infectee pairs observed. To accomodate small counts, the non-parameteric Monte Carlo version of the chi-square test (999 replicates) was then used \cite{pearson_x_1900,bradley_monte_1977}.
\added{This formulation assumes equal forest sizes ($m_A = m_B = m$). Under the null hypothesis that both forests are sampled from the same distribution of infector-infectee pairs, the expected count for pair $(i,j)$ in forest $\mathcal{F}_A$ is $E_{ij}^{\mathcal{F}_A} = \frac{c_{ij}^{\mathcal{F}_A} + c_{ij}^{\mathcal{F}_B}}{2}$, and similarly for $\mathcal{F}_B$. Substituting these expected values into the classical chi-squared formula $\frac{(O-E)^2}{E}$ and simplifying yields Equation \ref{eq:chisq}.}

\subsection{PERMANOVA}
PERMANOVA is a generic approach used to test group differences using pairwise distances between all observations of a sample \added{and makes no model assumptions} \cite{anderson_new_2001}. Here, we apply it to test whether distances between transmission trees differ when the trees belong to the same epidemic forest versus different forests. 
\subsubsection{Distance between two transmission trees}
The field of phylogenetics offers a range of established methods for comparing tree structures, providing several distance metrics for quantifying topological differences between pairs of phylogenies
\cite{jombart_treespace_2017, pavoine_testing_2008,robinson_comparison_1979, steel_distributions_1993, colijn_metric_2018, kendall_estimating_2018}.
These methods typically follow a two-step process: (i) convert trees into vectors of pairwise distances between all sampled taxa and (ii) compute Euclidean distances between these vectors.

A commonly used metric for the first step is the \textit{patristic} distance \cite{steel_distributions_1993}, defined as the sum of branch lengths on the path separating two taxa, reflecting the evolutionary distance between them. Adapting this concept to transmission trees, we define the \replaced{graph}{patristic} distance between cases (\textit{i.e.} vertices) $v_i$ and $v_j$ as the sum of edge weights along their connecting path on the undirected graph. Since all edges here have a weight of 1, this distance directly corresponds to the number of transmission events between cases, carrying clear epidemiological meaning. An illustration of \replaced{graph}{patristic} distances in a transmission tree is available in the supplementary material (Fig.\ref{fig:patristic}).

We denote $\pi(.)$ the function mapping a transmission tree $T$ of size $n$ into a vector of $\frac{\added{n}(n-1)}{2}$ \replaced{graph}{patristic} distances:
\begin{equation}
    \mathbf{d}_T = \pi(T)
\end{equation}

where $\mathbf{d}_T \in \mathbb{R}_+^{n(n-1)/2}$.

The dissimilarity between two trees $ T_k $ and $ T_l $ is then quantified by the Euclidean distance between the respective vectors of \replaced{graph}{patristic} distances, calculated as the norm:
\begin{equation}
D(T_k, T_l) = \|  \mathbf{d}_{T_k} -  \mathbf{d}_{T_l} \|
\end{equation}

This distance captures topological differences by evaluating how the relative positions of vertices, encoded as \replaced{graph}{patristic} distances, diverge between the two trees. If $ T_k $ and $ T_l $ have identical edge sets, their \replaced{graph}{patristic} distance matrices are equal, yielding $ D(T_k, T_l) = 0 $; otherwise, discrepancies in path lengths increase the distance.

\subsubsection{Outline of the method}
Given two epidemic forests, $\mathcal{F_A}$ and $\mathcal{F_B}$, each containing $m$ transmission trees, we apply PERMANOVA to test whether tree topologies differ significantly between forests. Broadly, the method partitions pairwise distances between all trees into within-group ($\text{SS}_W$) and between-group ($\text{SS}_B$) components \cite{anderson_new_2001}, based on pre-defined groups (here, the two forests). Statistical significance is assessed through permutation testing, where forest labels are randomly reassigned multiple times.

We define the combined epidemic forest as $\mathcal{F_{A \bigcup B}} = \mathcal{F_A} \bigcup \mathcal{F_B}$, containing all trees from $\mathcal{F}_A$ and $\mathcal{F}_B$. The total sum of squares, $\text{SS}_T$, representing the overall variance across all trees in $\mathcal{F_{A \bigcup B}}$, is:
\begin{equation}
\text{SS}_T = \frac{1}{2m} \sum_{k=1}^{2m} \sum_{l=1}^{2m} D(T^{A \bigcup B}_k, T^{A \bigcup B}_l)^2
\end{equation}
The double summation computes squared pairwise distances amongst the $2m$ trees in $\mathcal{F_{A \bigcup B}}$, which decomposes to:
\begin{equation}
\text{SS}_T = \frac{1}{2m} \left( \sum_{k=1}^{m} \sum_{l=1}^{m} D(T^A_k, T^A_l)^2 + \sum_{k=1}^{m} \sum_{l=1}^{m} D(T^B_k, T^B_l)^2 + 2\sum_{k=1}^{m} \sum_{l=1}^{m} D(T^A_k, T^B_l)^2 \right)
\end{equation}

The within-group sum of squares $ \text{SS}_W $ measures the variance within forests:
\begin{equation}
\text{SS}_W = \frac{1}{m} \left( \sum_{k=1}^{m} \sum_{l=1}^{m} D(T^A_k, T^A_l)^2 +  \sum_{k=1}^{m} \sum_{l=1}^{m} D(T^B_k, T^B_l)^2 \right)
\end{equation}
where each term sums the squared distances among all pairs within each forest, normalised by $m$. The between-group sum of squares ($ \text{SS}_B $), capturing variability between the forests, is:
\begin{equation}
\text{SS}_B = \text{SS}_T - \text{SS}_W
\end{equation}
The PERMANOVA test statistic \cite{anderson_new_2001} is:
\begin{equation}
F = \frac{\text{SS}_B}{\text{SS}_W / (2m - 2)}
\label{eq:f-statistic}
\end{equation}

The reference distribution of $F$ under the null hypothesis of no differences between groups is generated by a Monte Carlo procedure where forests labels are permuted a large numbe\added{r} of times (\textit{i.e.} 999 by default). \textit{p}-values are calculated as the proportion of permuted $ F $-values exceeding the observed $ F $ \cite{anderson_new_2001}.

\subsection{Simulation study}
\begin{figure}
    \centering
    \includegraphics[width=1\linewidth]{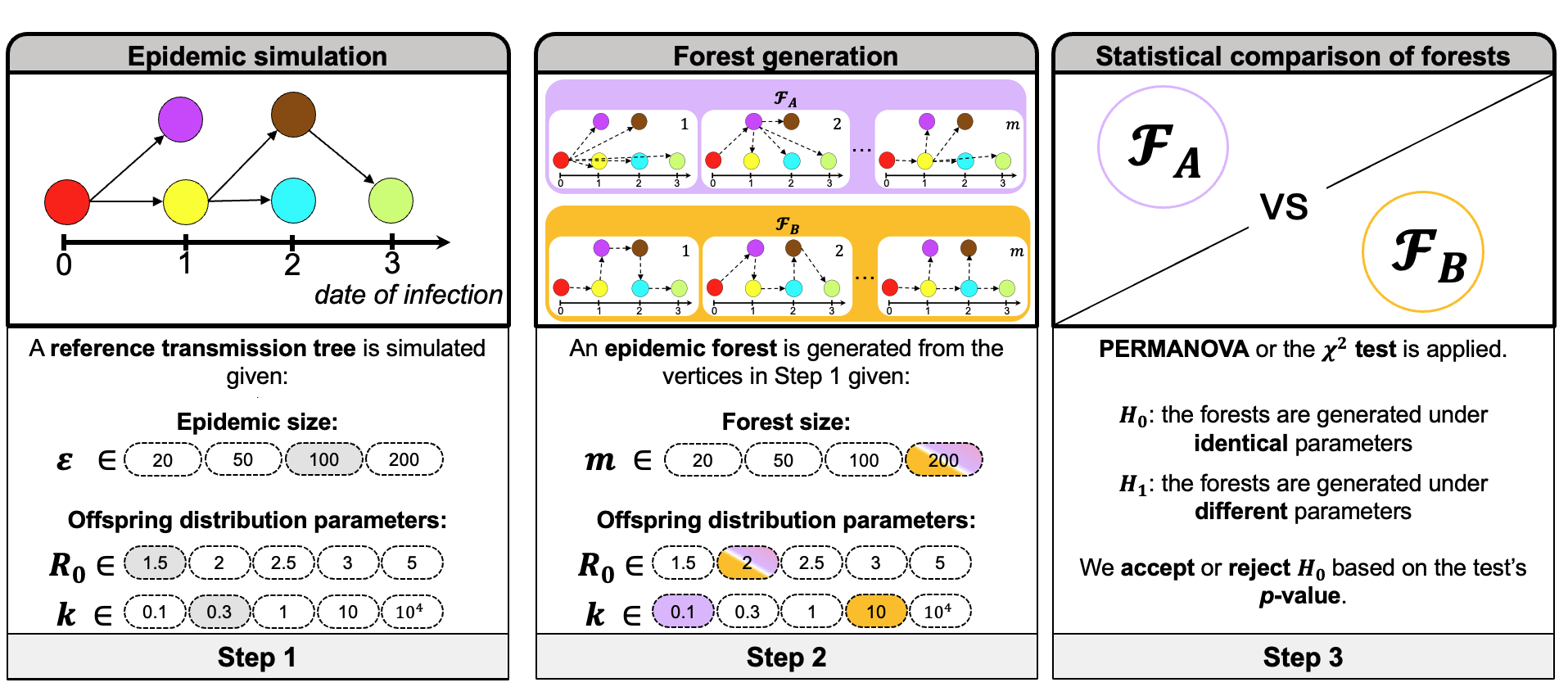}
      \caption{Simulation framework for assessing the performance of the $\chi^2$ test and PERMANOVA.}
    \medskip
    \small
    Diagram illustrating the simulation study to assess the respective performances of $\chi^2$ test and PERMANOVA for detecting differences between pairs of epidemic forests.
     \begin{enumerate}
         \item We simulate \replaced{a \textit{reference} transmission tree $\mathcal{T}$ with $\varepsilon$ infections from offspring distribution $\text{NegBin}(R_0, k)$}{`ground-truth’ transmission trees under two generative processes and pair them by epidemic size ($\epsilon$)}. This process is repeated 100 times to account for the stochasticity of epidemic dynamics.
         \item We generate reconstructed forests \replaced{$\mathcal{F}_A$ and $\mathcal{F}_B$, each containing $m$ trees, by re-assigning infector-infectee relationships from $\text{NegBin}(R_{0,A}', k_A')$ and $\text{NegBin}(R_{0,B}', k_B')$, conditional on $\mathcal{T}$'s dates of infection and case identifiers. In this example, $\mathcal{F}_A = \text{NegBin}(R_0 = 2, k = 0.1)$ and $\mathcal{F}_B = \text{NegBin}(R_0 = 2, k = 10)$}{($\mathcal{G}$) as the posterior samples of trees from \textit{outbreaker2} which takes as input some data from step 1, incorporating uncertainty on who infected whom}.
         \item \replaced{The $\chi^2$ test and PERMANOVA are applied to test whether the two epidemic forests stem from the same generative process.}{Epidemic forests ($\mathcal{F}$) are drawn from $\mathcal{G}$ under varying sampling conditions. Forest size ($m$) refers to the number of sampled trees, while overlap frequency ($\omega$) controls the similarity between forests by determining the probability that a tree in $\mathcal{F}_\text{Mix}$ is drawn from $\mathcal{G}_\text{NegBin}$. $\mathcal{F}_\text{NegBin}$ is strictly drawn from $\mathcal{G}_\text{NegBin}$.  This process is repeated 100 times to account for sampling stochasticity.}
     \end{enumerate}
    \label{fig:simulation}
\end{figure}

We conducted a simulation study to evaluate the performance of the $\chi^2$ test and PERMANOVA in distinguishing between simulated epidemic forests drawn from distinct generative processes corresponding to different epidemic dynamics. The simulation framework is illustrated in Figure \ref{fig:simulation}.

\clearpage
\subsubsection{Simulating epidemic forests}
\deleted{We simulated outbreaks under two contrasting generative processes, that give rise to different epidemic dynamics. Epidemics were modelled using a branching process (see supplementary information), where infections propagate through successive generations according to a specified offspring distributions.\\}
\added{We generated epidemic forests through a three-stage process to systematically evaluate forest comparison methods across diverse transmission scenarios.\\
First, we defined the parameter space for the simulations:} 
\begin{itemize}
    \item \added{\textbf{Epidemic size}: $\varepsilon \in\{20, 50, 100, 200\}$. The number of infected individuals, corresponding to the number of vertices in the tree.}
    \item \added{\textbf{Basic Reproduction Number}: $R_0\in\{1.5,2,3\}$. The mean number of secondary infections per case in a fully susceptible population, corresponding to the mean of the negative binomial offspring distribution.}
    \item \added{\textbf{Dispersion Parameter}: $k\in\{0.1,0.3,0.5,1,\infty\}$. Controls heterogeneity in individual transmission, corresponding to the dispersion of the negative binomial offspring distribution. Lower values indicate greater overdispersion; as $k \rightarrow \infty$, the distribution converges to Poisson.}
\end{itemize}
\added{For each epidemic sizes $\varepsilon$, we defined offspring distributions $\text{NegBin}(R_0, k)$ using all pairwise combinations of basic reproduction number $R_0$ and dispersion parameter $k$.}\\
\added{Second, we generated \textit{reference} transmission trees. For each parameter combination $(\varepsilon, R_0, k)$ we simulated a reference transmission tree $\mathcal{T}^{\varepsilon}_{(R_0, k)}$ using a stochastic branching process. Secondary infections per case were drawn from $\text{NegBin}(R_0, k)$ , and generation times followed a gamma distribution with mean of 12 days and standard deviation of 6 days. We generated 100 replicate trees per parameter set to account for stochasticity. Simulations were initialised with 10,000 susceptible individuals, ran for a maximum of 365 days, and terminated upon reaching exactly $\varepsilon$ infections, thereby excluding saturation effects. Within each reference tree $\mathcal{T}^{\varepsilon}_{(R_0, k)}$, infected individuals were assigned identifiers $v \in \{1, \ldots, \varepsilon\}$, ordered by their dates of infection $t_v$.\\
Third, we constructed epidemic forests by re-assigning cases' ancestries. For each reference tree $\mathcal{T}^{\varepsilon}_{(R_0, k)}$, we generated forests $\mathcal{F}_{\mathcal{T}^{\varepsilon}_{(R_0, k)}}(R_0', k')$ by conditioning on the observed infection set $\mathcal{I}_{\mathcal{T}^{\varepsilon}_{(R_0, k)}} = \{(v, t_v)\}_{v=1}^{\varepsilon}$ while resampling ancestries from $\text{NegBin}(R_0', k')$. Each forest comprised of $m=200$ trees. This procedure yielded 15 distinct forests per reference tree (one for each offspring distribution pair $(R_0', k')$), including one forest matched the reference tree's generative process, where $(R_0', k') = (R_0, k)$ .\\
This procedure generated a total of 6,000 reference trees ($|\varepsilon| \times |R_0| \times |k| \times \text{ replicates} = 4 \times 3 \times 5 \times 100$), each generating 15 distinct forests ($ |R_0'| \times |k'|$), yielding 120 pairwise forest comparisons per reference tree ($\binom{15}{2} + 15$), resulting in a total of 720,000 forest comparisons.}\\

\deleted{For each generative process, we simulated 100 outbreaks for each epidemic size category ($ \epsilon \in \{20, 50, 100, 200\} $), allowing a $ \pm 20\% $ tolerance on the number of cases (\textit{i.e.} vertices) due to stochasticity (see supplementary information). For a given epidemic size category, we paired simulated transmission trees with matching vertex count from both generative processes (step 1 in Fig.\ref{fig:simulation}). Each simulation produced a ground-truth transmission tree and symptom onset dates for all cases (see supplementary information), which served as input data for subsequent Bayesian inference of who infected whom.}\hfill \break

\deleted{For each simulated outbreak, we obtained 180 posterior samples of transmission trees derived from oubtreak reconstruction, namely $\mathcal{G}_\text{Pois}$ and $\mathcal{G}_\text{NegBin}$ (step 2 in Fig.\ref{fig:simulation}). Given the absence of simulated genetic data and to ensure sufficient reconstruction accuracy, we incorporated partial, randomly selected, contact tracing information into the inference process \cite{campbell_bayesian_2019} — 50\% of true infectious contacts for the superspreading scenario and 65\% for the homogeneous scenario (see supplementary information for details on the \textit{outbreaker2} analyses).}

\deleted{To consider a wide range of contexts, we varied three parameters in our simulation study (dashed elements in Fig.\ref{fig:simulation}):}

\subsubsection{Assessing statistical performance}
\replaced{For each of the 720,000 forest comparisons ($\mathcal{F}_A$ vs. $\mathcal{F}_B$), we performed the $\chi^2$ test and PERMANOVA under 4 forest sizes $m \in {20, 50, 100, 200}$, where $m$ denotes denotes the number of trees sampled from each forest. This resulted in a total of 5,760,000 tests performed.}{A total of 960,000 tests (96 parameter combinations, 100 simulated epidemic pairs, 100 sampling replicates) were conducted, using both the $\chi^2$ approach and the PERMANOVA.}
For each parameter combination, we measured:
\begin{itemize}
    \item \textbf{Sensitivity}: The proportion of tests that correctly rejected the null hypothesis ($H_0$) when comparing \replaced{forests generated with different offspring distributions, \textit{i.e.}, $\mathcal{F}_{\mathcal{T}}(R_0, k)$ vs. $\mathcal{F}_{\mathcal{T}}(R_0', k')$ where $(R_0, k) \neq (R_0', k')$}{(partially) distinct epidemic forests ($\omega < 1$)}.
    \item \textbf{Specificity}: The proportion of tests that correctly accepted $H_0$ when comparing forests \replaced{generated with}{drawn from} identical \added{offspring} distributions \replaced{, \textit{i.e.}, $(R_0, k) = (R_0', k')$}{($\omega = 1$)}.
    \item \deleted{\textbf{Area Under the Cure (AUC)}: ROC curves were derived (Fig. \ref{fig:roc-curves}) and AUC were calculated as a general indicator of the test's performance. Values of AUC = 1 indicate perfect sensitivity and specificty.}
\end{itemize}

\added{To quantify the factors influencing test sensitivity, we fit a logistic regression model to all comparisons where forests were generated under different parameter settings ($H_1$; $n=5{,}040{,}000$). The binary outcome was whether the test correctly rejected the null hypothesis ($H_0$). We compared four nested models using the Akaike Information Criterion and selected the model with the lowest value. The final model included main effects for statistical method (PERMANOVA or $\chi^2$), forest size ($m$), epidemic size ($\varepsilon$), and parameter differences between forests ($\Delta R_0$ and $\Delta k$). It also included all two way and three way interaction terms involving the method:}

\begin{equation}
\begin{split}
\text{logit}(P(\text{reject } H_0)) &= 
\beta_0 + \beta_{\text{method}} + \beta_{m} + \beta_{\varepsilon} 
+ \beta_{\Delta R_0} + \beta_{\Delta k} \\
&\quad + \beta_{\text{method} : \varepsilon} 
+ \beta_{\text{method} : \Delta R_0} 
+ \beta_{\text{method} : \Delta k} \\
&\quad + \beta_{\Delta R_0 : \Delta k} 
+ \beta_{\text{method} : \Delta R_0 : \Delta k} + e
\end{split}
 \label{eq:sensitivity}
\end{equation}
\added{Where `:' represent the interaction term and $e$ is the normally distributed residuals. Results are reported as odds ratios using the $\chi^2$ test, the smallest forest size ($m=20$), the smallest epidemic size ($\varepsilon=20$), and no difference in dispersion ($\Delta k = 0$) as reference categories. The model achieved a pseudo $R^2$ of 0.58 \cite{tjur_coefficients_2009}.\\
Both methods achieved near-perfect specificity ($>97\%$) across all conditions, precluding regression analysis.}

\clearpage

\section{Author contributions}
\begin{itemize}
  \item Conceptualisation: CG, AC, TJ
  \item Methodology: CG, AC, TJ
  \item Software: CG
  \item Validation: CG, AC, TJ
  \item Visualization: CG
  \item Writing – original draft: CG
  \item Writing – review \& editing: CG, AC, TJ
  \item Supervision: AC, PJW, TJ
  \item Funding acquisition: PJW
\end{itemize}

\section{Funding and competing interests}
This work was funded by the National Institute for Health and Care Research (NIHR) Health Protection Research Unit (HPRU) in Modelling and Health Economics, which was a partnership between Imperial College London, London School of Hygiene and Tropical Medicine, and UKHSA (grant code NIHR200908). All authors declare that they have no competing interests.

\section{Data and materials availability}
Simulations, analyses and visualisations were performed using the R software version 4.4.0 (https://www.R-project.org/) \cite{r_core_team_r_2025}. 
Our framework has been implemented in a free, open-source, R package \textit{mixtree}, which is available on CRAN \cite{geismar_mixtree_2025}. This study is fully reproducible using code available on GitHub: https://github.com/CyGei/mixtree\_analysis \added{(archived on Zenodo:10.5281/zenodo.17704758 \cite{cygei_cygeimixtree_analysis_2025}). The resulting data is stored on a Zenodo archive: 10.5281/zenodo.17704455 \cite{geismar_mixtree-analysis-data_2025}}.
\clearpage
\printbibliography
\clearpage
\input{supplementary}
\clearpage
% Add references cited only in the supplement
%\nocite{steel_distributions_1993,
%        jombart_bayesian_2014,
%        campbell_outbreaker2_2018,
%        lloyd-smith_superspreading_2005,
%        vink_serial_2014,
%        rai_estimates_2021,
%        geismar_bayesian_2023}

\end{document}

%% file: supplementary.tex
\renewcommand{\thefigure}{S\arabic{figure}}
\setcounter{figure}{0}  % Reset figure counter for supplementary
\renewcommand{\thesection}{S\arabic{section}}
\setcounter{section}{0}  % Reset section counter for supplementary
\renewcommand{\thetable}{S\arabic{table}}
\setcounter{table}{0}  % Reset table counter for supplementary material

\section*{Supplementary Information}

The following presents the supplementary materials for the paper entitled: `\textit{A statistical framework for comparing epidemic forests}'. The first part illustrates the computation of \replaced{graph}{patristic} distances in transmission trees, the second details the methodology for simulating epidemic forests, and the third presents additional results.

\section{Graph distances}
\deleted{The \textit{patristic} distance, originally used for evolutionary trees \cite{steel_distributions_1993}, is defined as the sum of branch lengths along the path separating two taxa. Applied to transmission trees} We define the \replaced{graph}{patristic} distance between cases as the number of undirected edges along their connecting path, corresponding to the number of transmission events between the considered cases.
The diagram  below illustrates this concept using a transmission tree (panel A) with five cases. Panel B shows the corresponding matrix of \replaced{graph}{patristic} distances between all pairs of cases. Distances between case 5 and all other cases are highlighted in colour, corresponding to the coloured paths in panel A. This process is conducted in step 1 of the right panel in Figure \ref{fig:diagram}  of the main text.
\begin{figure}[!h]
    \centering
    \includegraphics[width=1\linewidth]{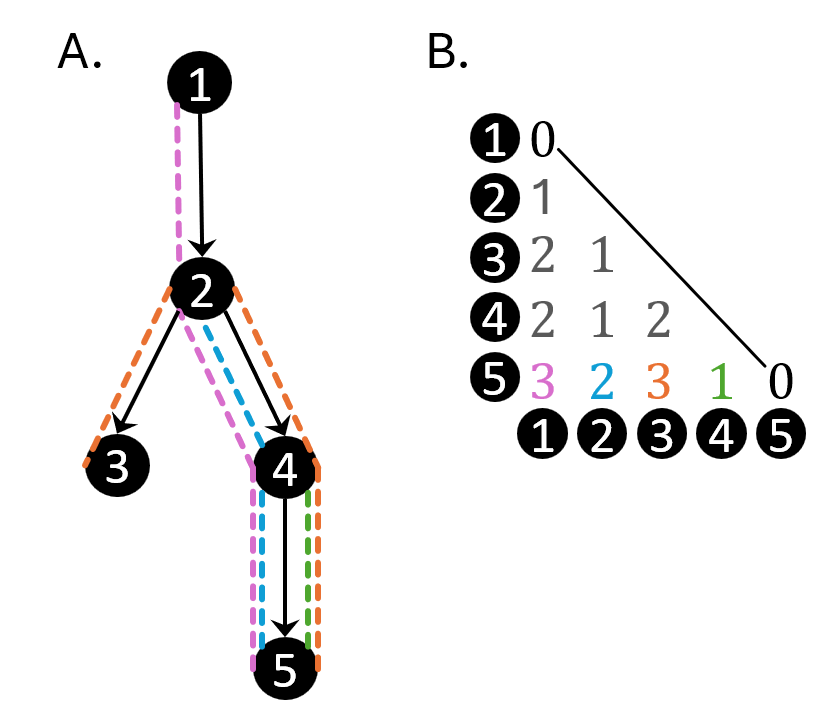}
    \caption{Calculation of \replaced{graph}{patristic} distances in a transmission tree.}
    \medskip
    \small
    A. A transmission tree with 5 cases. The coloured dashed lines show the unique paths connecting case 5 to all other cases.
    B.   The matrix representation of \replaced{graph}{patristic} distances between all pairs of cases. The coloured numbers correspond to the number of transmission events between case 5 and all other cases, matching the coloured paths shown in A. For illustration, we only represent the lower triangle but the matrix is symmetric.
    \label{fig:patristic}
\end{figure}
\clearpage
\section{Simulation Framework}
Our simulation study followed a two-step process designed to evaluate how effectively the proposed statistical methods could distinguish between epidemic forests derived from distinct generative processes. First, we simulated \replaced{\textit{reference}}{ground-truth} transmission trees using \replaced{a branching process model}{the \textit{simulacr} R package (https://github.com/CyGei/simulacr), generating outbreaks with} \replaced{with varying offspring distributions}{and without superspreading events}. Second, we generated epidemic forests from these \replaced{reference}{ground-truth} trees \replaced{by re-assigning infector-infectee relationships from a new offspring distribution, conditional on the reference's tree case identifiers and dates of infection}{using Bayesian inference with \textit{outbreaker2} \cite{jombart_bayesian_2014, campbell_outbreaker2_2018}}, \deleted{which produced} \added{producing} collections of plausible transmission trees that reflect the underlying dynamics of each scenario. This approach provided a controlled environment with known ground truth against which we could systematically assess the discriminatory power of our statistical tests across different epidemic sizes, \replaced{forest}{sample} sizes, and \replaced{offspring distribution parameters}{degrees of similarity between compared forests (\textit{overlap frequency})}.

\subsection{Branching process model}
Our simulations employ the discrete-time stochastic branching process implemented in \textit{simulacr} (https://github.com/CyGei/simulacr). Each simulation was initiated with a single infected individual (the index case), from whom infections propagated across successive generations based on specified offspring and generation time distributions.\hfill \break

\textit{simulacr} tracks the propagation of infections through successive generations by computing the force of infection (FOI) at each time step $[t,t+1)$. For each infected individual $i$, we define $t_i$ as their infection time and $R_i$ as their case reproduction number—the expected number of secondary cases they generate in a fully susceptible population. The generation time—the interval between the infection of a primary case and the infection of its secondary cases—follows a probability mass function $g(t)$, where $g(t)=0$ for $t \leq 0$. At each time step, the model sums the infectious contribution from all active cases to determine the overall force of infection, then probabilistically generates new infections from the susceptible population ($S(t)$) and assigns their respective infectors according to their relative contribution to transmission.\hfill \break

The FOI generated by case $i$ at time $t$ is defined as:
\begin{equation}
\lambda_i(t) = R_i\,g(t-t_i)
\end{equation}
The total FOI at time $t$ arising from all infectious individuals at this time (denoted by the set $I(t)$) is given by:
\begin{equation}
\Lambda(t) = \sum_{i \in I(t)} \lambda_i(t) 
\end{equation}
Each susceptible individual $j$ (with $j\in S(t)$) faces an infection probability during the interval $[t,t+1)$ of:
\begin{equation}
p_j = 1 - e^{-\Lambda(t)}
\end{equation}

Once a susceptible individual becomes infected at time $t+1$, a specific infector is drawn from a multinomial distribution where the probability that case $i \in I(t)$ is the infector is defined as: 
\begin{equation}
P(i \mid \text{infection at } t+1) = \frac{\lambda_i(t)}{\Lambda(t)}
\end{equation}
\replaced{To achieve exactly $\varepsilon$ cases per outbreak, we initialised simulations with 10,000 susceptible individuals and terminated them upon reaching exactly $\varepsilon$ infections, thereby excluding saturation effects. This truncation scheme led to right-censoring, where cases infected within one generation time of the truncation date likely had not realised their expected number of secondary infections $R_i$.}{This process iterates until the epidemic extinguishes or the maximum number of days (\textit{i.e.} 100) has been reached.} \hfill \break

\begin{figure}[!h]
    \centering
    \includegraphics[width=1\linewidth]{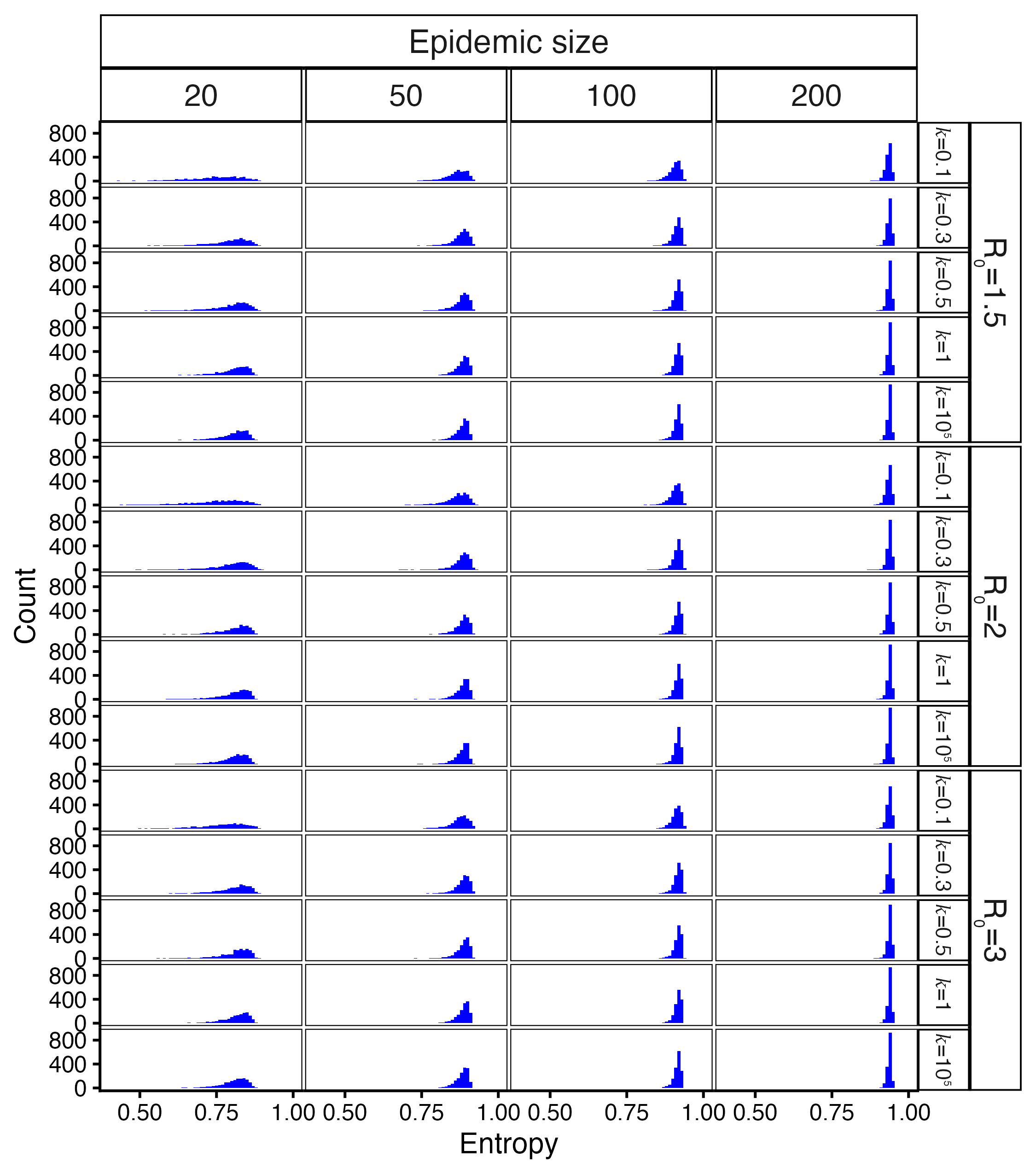}
    \caption{Variation in infector-infectee relationships across epidemic forests}
    \medskip
    \small
    \justifying
    \added{Histogram of mean scaled entropy across epidemic forests ($m = 200$ trees per forest), stratified by simulation parameters. For each infectee $j$, the scaled entropy $H_j$ (x axis) quantifies variation in their assigned infector across all trees in a forest, computed using the normalised Shannon entropy formula \cite{shannon_mathematical_1948}: 
    $H_j= \frac{-\sum_{i=1}^{K_j} p_{ij} \log(p_{ij})}{\log(K_j)}$, where $p_{ij}$ is the proportion of trees in which individual $i$ infects $j$, and $K_j$ is the number of distinct infectors of $j$ observed across the forest.  Values range from 0 (identical infector in all trees) to 1 (all possible infectors equally frequent). The mean scaled entropy ($\bar{H}$) is obtained by averaging $H_j$ over all cases. Columns refer to epidemic sizes ($\varepsilon$), rows refer to the mean reproduction number $R_0$ and dispersion parameter $k$ of the negative binomial offspring distribution. Average entropy for our simulations is 77\% and increases with epidemic size, due to greater variation in infector assignment (Table\ref{tab:lm-entropy}).
    \label{fig:entropy.png}}
\end{figure}

For \replaced{all simulations}{both epidemic scenarios}, we modelled the generation time using a discretised probability mass function (PMF) with a mean of \replaced{12}{4} days and a standard deviation of \replaced{6}{0.84} days \added{to enable considerable entropy across reconstructed forests (supplementary Fig. \ref{fig:entropy.png})}. \deleted{This parameterisation is consistent with estimates for common respiratory pathogens including influenza and SARS-CoV-2 \cite{vink_serial_2014,rai_estimates_2021,geismar_bayesian_2023}. We applied the same distribution to model the incubation period (time from infection to symptom onset).}

Each simulated transmission tree returns the case identifiers of infected individuals and their \replaced{infection}{symptom onset} dates which will be used for \replaced{forest generation}{outbreak reconstruction}.

\subsection{\replaced{Forest generation}{Outbreak reconstruction}}
For each simulated \added{reference transmission tree  $\mathcal{T}^{\varepsilon}_{(R_0, k)}$, we generated epidemic forests $\mathcal{F}_{\mathcal{T}^{\varepsilon}_{(R_0, k)}}$ by reassigning infector-infectee relationships given a new offspring distribution $\text{NegBin}(R_0', k')$ while conditioning on the reference tree's set of infection $\mathcal{I}_{\mathcal{T}^{\varepsilon}_{(R_0, k)}} = \{(v, t_v)\}_{v=1}^{\varepsilon}$, where $v$ denotes case idenitifers and $t_v$ their infection times. Each forest comprised of $m$ = 200 trees.} \deleted{outbreak, we reconstructed the posterior distribution of transmission trees using \textit{outbreaker2} \cite{jombart_bayesian_2014, campbell_outbreaker2_2018}, a Bayesian inference framework for outbreak reconstruction. \textit{outbreaker2} integrates information on the dates of symptom onset of cases, generation time and contact tracing data to probabilistically reconstruct transmission trees. Although the model supports genetic data, we did not simulate pathogen genome sequences as part of this simulation study.}\hfill \break
\added{For each tree in the forest, we followed a three-step process:}
\begin{itemize}
    \item \added{\textbf{Offspring sampling:} we drew individual reproduction numbers $R_i \sim \text{NegBin}(R_0', k')$ for all $\varepsilon$ cases. To ensure that the epidemic sets off, we constrained the index (root) case to have $R_1 \geq 1$ by sampling from a truncated negative binomial distribution when $R_1 < 1$.}
    \item \added{\textbf{Force of infection (FOI) calculation:} For each case $i$, with infection time $t_i$ and reproduction number $R_i$, we computed the FOI exerted by $i$ at time $t$ as:}
    \begin{equation*}
        \added{\lambda_i(t) = R_i g(t - t_i)}
    \end{equation*}
   \added{where $g()$ is the generation time probability mass function.}
    \item \textbf{Ancestry assignment:} cases were assigned infectors sequentially in order of their infection times. For each case $j$, infected at time $t_j$, we identified the set of ancestors $\mathcal{A}(t_j) = {i : t_i < t_j}$ and selected the infector by sampling from the multinomial distribution with probability relative to the contribution of $i$ on the whole FOI at time $t_j$:
    \begin{equation*}
    \added{P(i \rightarrow j) = \frac{\lambda_i(t_j)}{\sum_{i' \in \mathcal{A}(t_j)} \lambda_{i'}(t_j)}}
    \end{equation*}
    \added{This procedure ensures that each generated tree $\mathcal{T}_k = (V, E_k)$ in forest $F_{\mathcal{T}}(R_0', k')$ shares the same vertex set $V$ and infections times ${t_v}$ as the reference tree $\mathcal{T}$, but with edges resampled according to the FOI generated under the new offspring distribution $\text{NegBin}(R_0', k')$. An epidemic forest is thus a collection of transmission trees that share the same set of infected individuals but different ancestral relationships, all consistent with a given offspring distribution.}
\end{itemize}

\deleted{In the absence of genetic data, we incorporated partial contact tracing information randomly sampled from ground-truth transmission trees to ensure sufficient reconstruction accuracy (\textit{i.e.} the proportion of correctly assigned ancestries). Through preliminary calibration experiments, we determined that including 50\% of true infectious contacts for the superspreading scenario ($A$) and 65\% for the non-overdispersed scenario ($B$) consistently yielded reconstruction accuracy between 60-80\% across epidemic sizes.}\hfill \break

\deleted{We fixed the number of introduction, the reporting parameter ($\pi$) and the number of generations separating cases ($\kappa$) to 1, ensuring that all reconstructed transmission trees have no unobserved cases and a single introduction. For each inference, we used identical generation time and incubation period distributions from the simulation model. The MCMC was executed for 10,000 iterations, with a thinning of 1 in 50 and a burn-in of 1000 iterations, resulting in a total of 180 posterior samples.}
\clearpage
\section{Additional Results}

\begin{figure}[!h]
    \centering
    \includegraphics[width=1\linewidth]{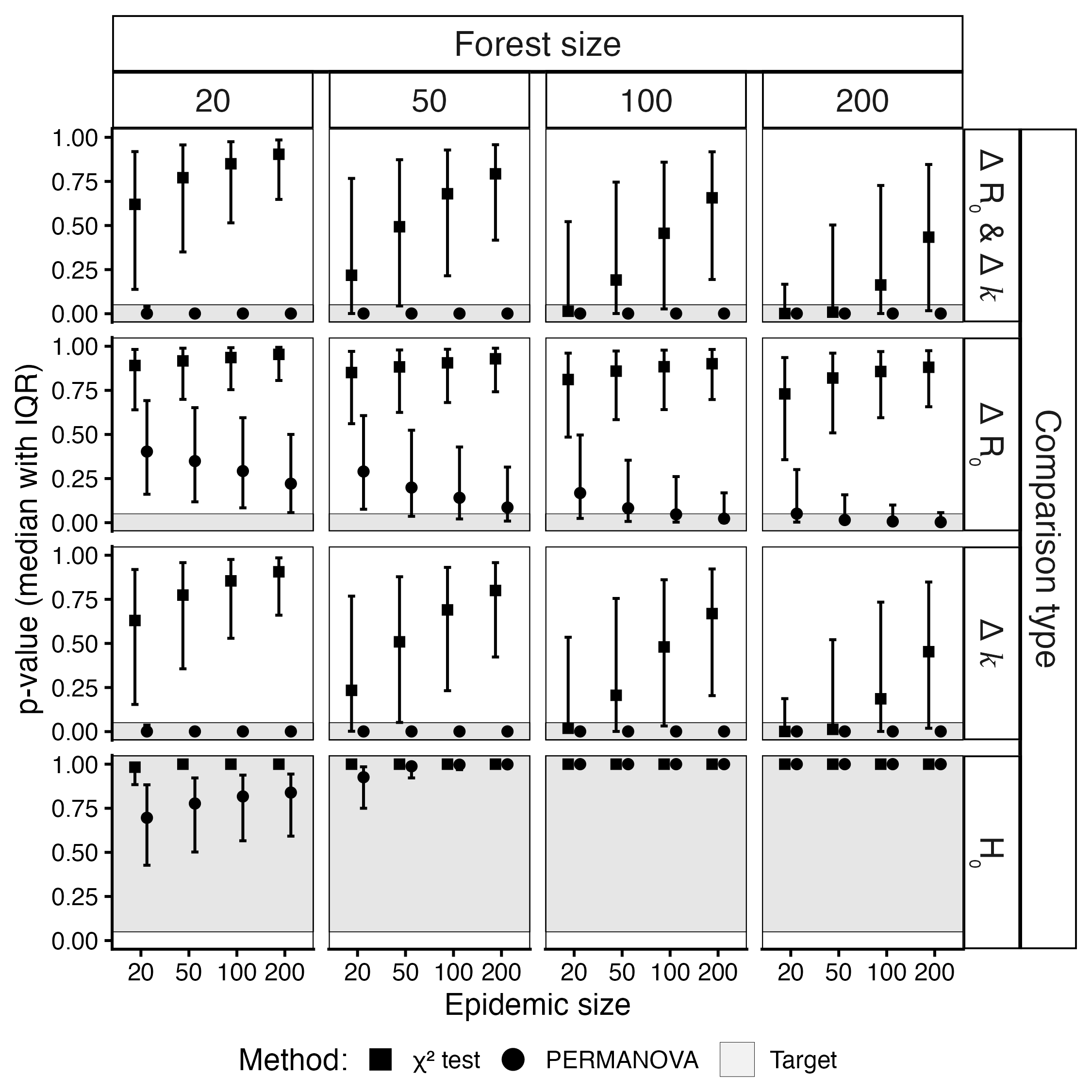}
    \caption{Performance of $\chi^2$ test and PERMANOVA for distinguishing epidemic forests.}
    \medskip
    \small
    \justifying
     Median p-values and interquartile ranges for the $\chi^2$ test (squares) and PERMANOVA (circles) across epidemic sizes (x-axis), forest sizes (columns), and parameter conditions (rows). Grey shading indicates desired p-value ranges: below $\alpha = 0.05$ when forests differ in at least one parameter (rows 1–3, reject $H_0$) and above $\alpha = 0.05$ when forests share identical parameters (row 4, accept $H_0$).
    \label{fig:delta_pval.png}
\end{figure}

\begin{figure}[!h]
    \centering
    \includegraphics[width=1\linewidth]{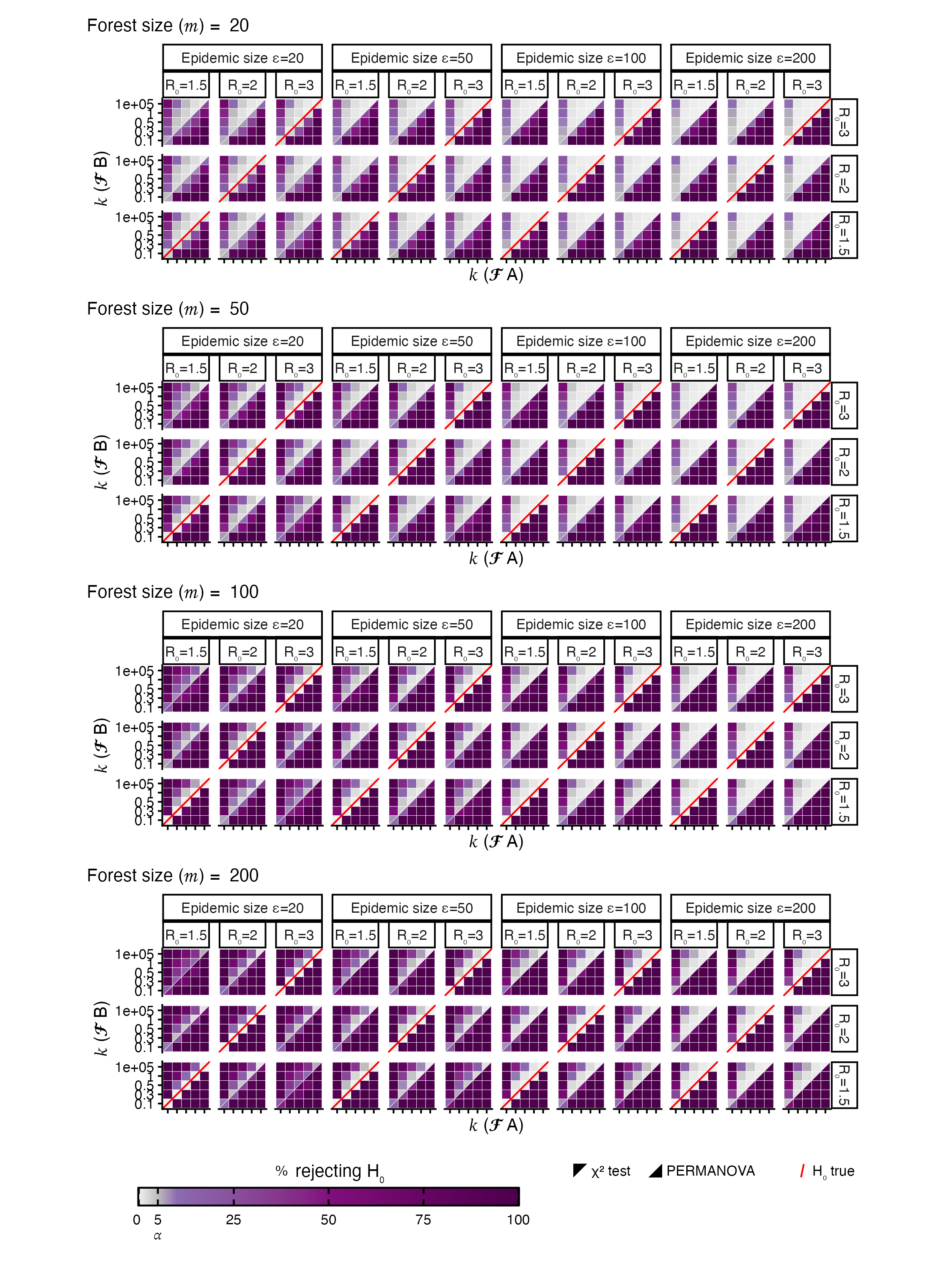}
    \caption{Performance of $\chi^2$ and PERMANOVA in distinguishing epidemic forests}
    \medskip
    \small
    \justifying
  \added{Each panel shows the proportion of tests rejecting the null hypothesis (p $<$ 0.05) when comparing epidemic forest $\mathcal{F_A}$ and $\mathcal{F_B}$. The upper triangle shows the $\chi^2$ test results; the lower triangle shows PERMANOVA results. Outer columns refer to epidemic size ($\varepsilon$), common to both forest. Forests can differ in their offspring distribution parameter: $R_0$ (inner columns) and $k$ (x and y axes; x-axis labels omitted for clarity, values identical to the y-axis). Red diagonal lines indicate comparisons where both forests share identical parameters ($H_0$ true; low rejection rates indicate good specificity). The other cells compare forests with different parameters (high rejection rates indicate good sensitivity). Both methods maintain excellent specificity (diagonal), but PERMANOVA demonstrates superior sensitivity.}
    \label{fig:grid_combined}
\end{figure}

\begin{figure}[!h]
    \centering
    \includegraphics[width=1\linewidth]{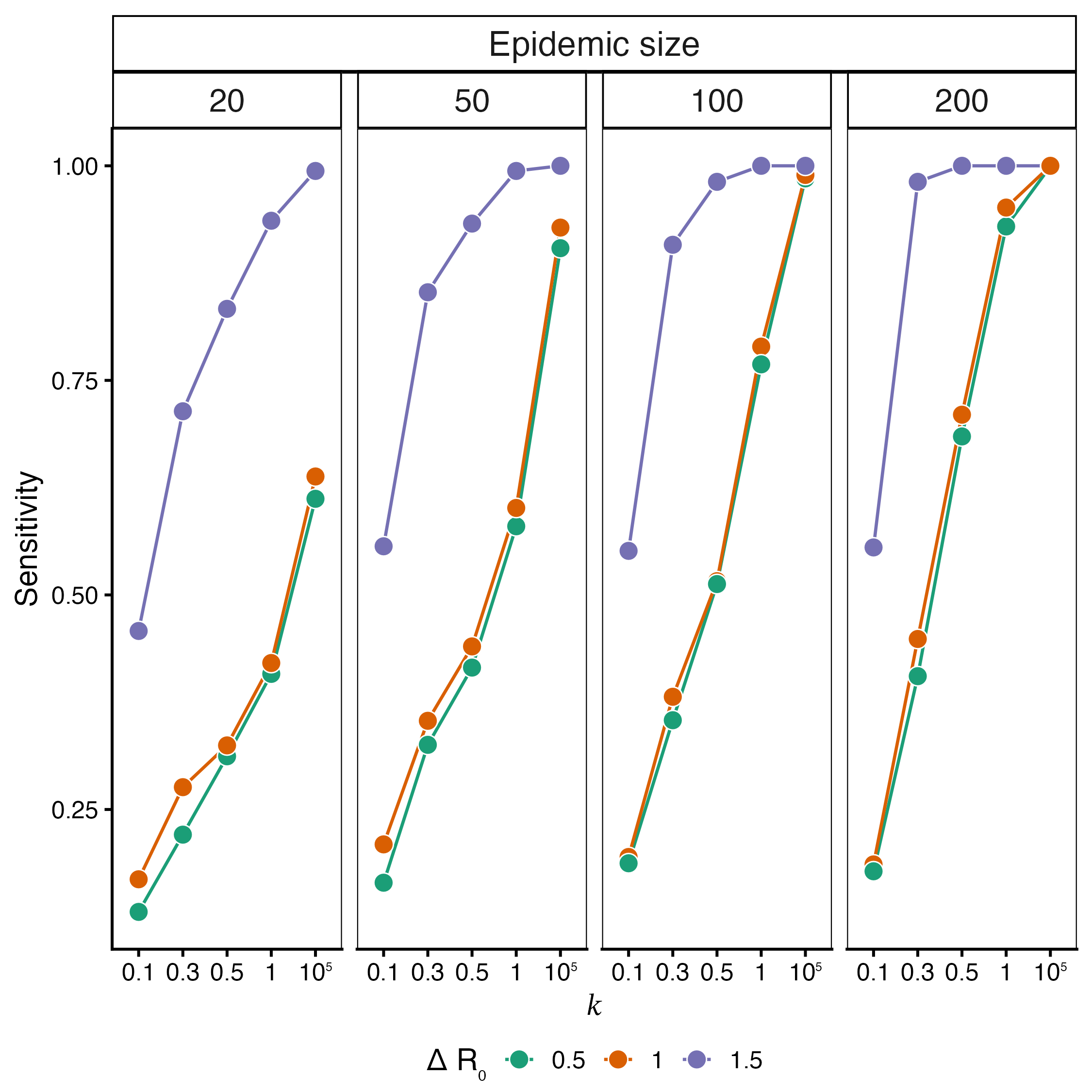}
    \caption{Sensitivity of PERMANOVA when comparing epidemic forests that differ only in $R_0$.}
    \medskip
    \small
    \justifying
    \added{Sensitivity (y-axis) is the proportion of tests correctly rejecting the null hypothesis when comparing forests of 200 trees generated with different $R_0$ ($\Delta R_0$, colour) but identical $k$ (x-axis). Columns correspond to epidemic size.}
    \label{fig:delta_R0_lineplots_m200}
\end{figure}
\clearpage

\clearpage
\added{The mean scaled entropy $\bar{H}$ for each forest was modelled as a linear function of epidemic size, reproduction number, and dispersion:}
\begin{equation}
    \bar{H^*} = \beta_0 + \beta_{\varepsilon} + \beta_{R_0} + \beta_k + \epsilon
    \label{eq:entropy}
\end{equation}
\added{where $\beta_0$ is the intercept, each $\beta$ term represents the categorical effect of epidemic size ($\varepsilon$), reproduction number ($R_0$), and dispersion parameter ($k$) respectively, and $\epsilon$ is the residual error. The model explained 68.2\% of the variance ($R^2 = 0.682$) in mean scaled entropy across 90,000 simulated forests, with coefficient estimates shown in Table \ref{tab:lm-entropy}.}

\begin{table}[h]
\centering
\caption{Linear regression results for the mean scaled entropy ($\bar{H}$) model (Eq. \ref{eq:entropy})}
\label{tab:lm-entropy}
\begin{tabular*}{\linewidth}{@{\extracolsep{\fill}} l l S[table-format=2.3] S[table-format=<1.3] }
\toprule
\multicolumn{2}{l}{\textbf{Predictor}} & {\text{\textbf{Estimate}}} & {\textbf{$p\text{-value}$}} \\
\midrule
Intercept & & 0.771 & {$<$0.001} \\
\addlinespace
\multirow{3}{*}{\textbf{Epidemic size ($\varepsilon$)}} 
 & 50 & 0.087 & {$<$0.001} \\
 & 100 & 0.123 & {$<$0.001} \\
 & 200 & 0.147 & {$<$0.001} \\
\addlinespace
\multirow{2}{*}{\textbf{Reproduction number ($R$)}} 
 & 2 & 0.003 & {$<$0.001} \\
 & 3 & 0.006 & {$<$0.001} \\
\addlinespace
\multirow{4}{*}{\textbf{Dispersion ($k$)}} 
 & 0.3 & 0.018 & {$<$0.001} \\
 & 0.5 & 0.020 & {$<$0.001} \\
 & 1 & 0.021 & {$<$0.001} \\
 & Poisson & 0.021 & {$<$0.001} \\
\addlinespace
\bottomrule
\multicolumn{4}{p{0.9\linewidth}}{\footnotesize \textit{Note:} Model fit $R^2 = 0.682$. All p-values are $<0.001$ due to the large simulation sample size ($n = 90{,}000$). Coefficient estimates indicate the expected change in mean scaled entropy relative to the reference category ($\varepsilon = 20$, $R = 1.5$, $k = 0.1$).}
\end{tabular*}
\end{table}

\clearpage

\begin{figure}[!h]
    \centering
    \includegraphics[width=1\linewidth]{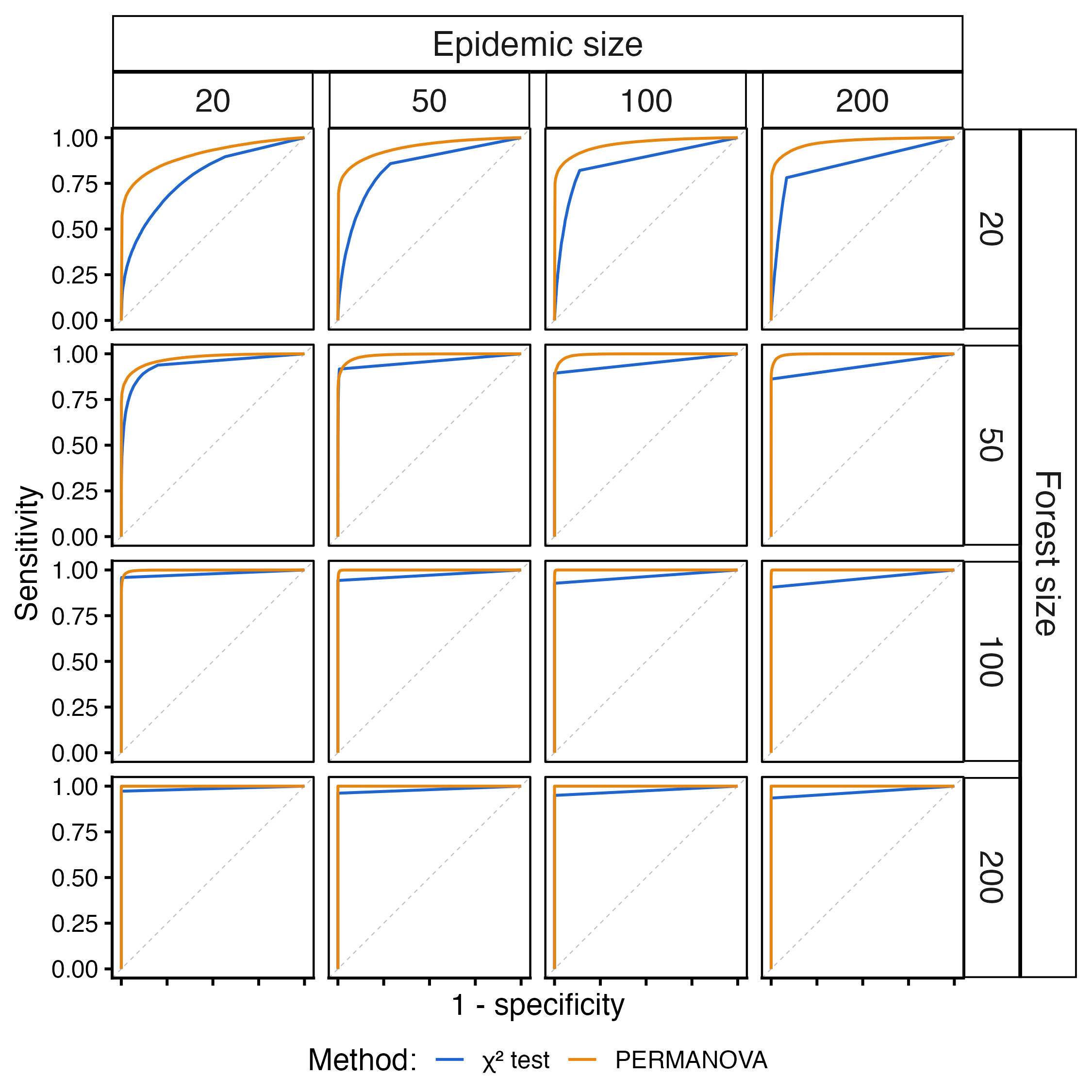}
    \caption{ROC curves for the $\chi^2$ test and PERMANOVA}
    \medskip
    \small
    \justifying
    Each panel shows the receiver operating characteristic (ROC) curves plotting true positive rate (sensitivity) against false positive rate (1-specificity) for the $\chi^2$ test \replaced{(blue)}{(green)} and PERMANOVA \replaced{(orange)}{(red)} across all \added{simulations for all} possible significance thresholds ($0 \leq \alpha \leq 1$).
    Panels are arranged by epidemic size (columns: 20–200 cases) and forest size (rows: 20–200 trees), x-axis tick labels are omitted for clarity, as both axes share the same scale.\deleted{, with all panels zoomed to the upper-left quadrant (FPR: 0-0.3, TPR: 0.7-1) for better visualisation; the full scale is shown in the external left panel for illustration. Points refer to specific $\alpha$ values (0.0015, 0.01, 0.05, 0.10).} 
    \label{fig:roc-curves}
\end{figure}
\clearpage
\begin{figure}[!h]
    \centering
    \includegraphics[width=1\linewidth]{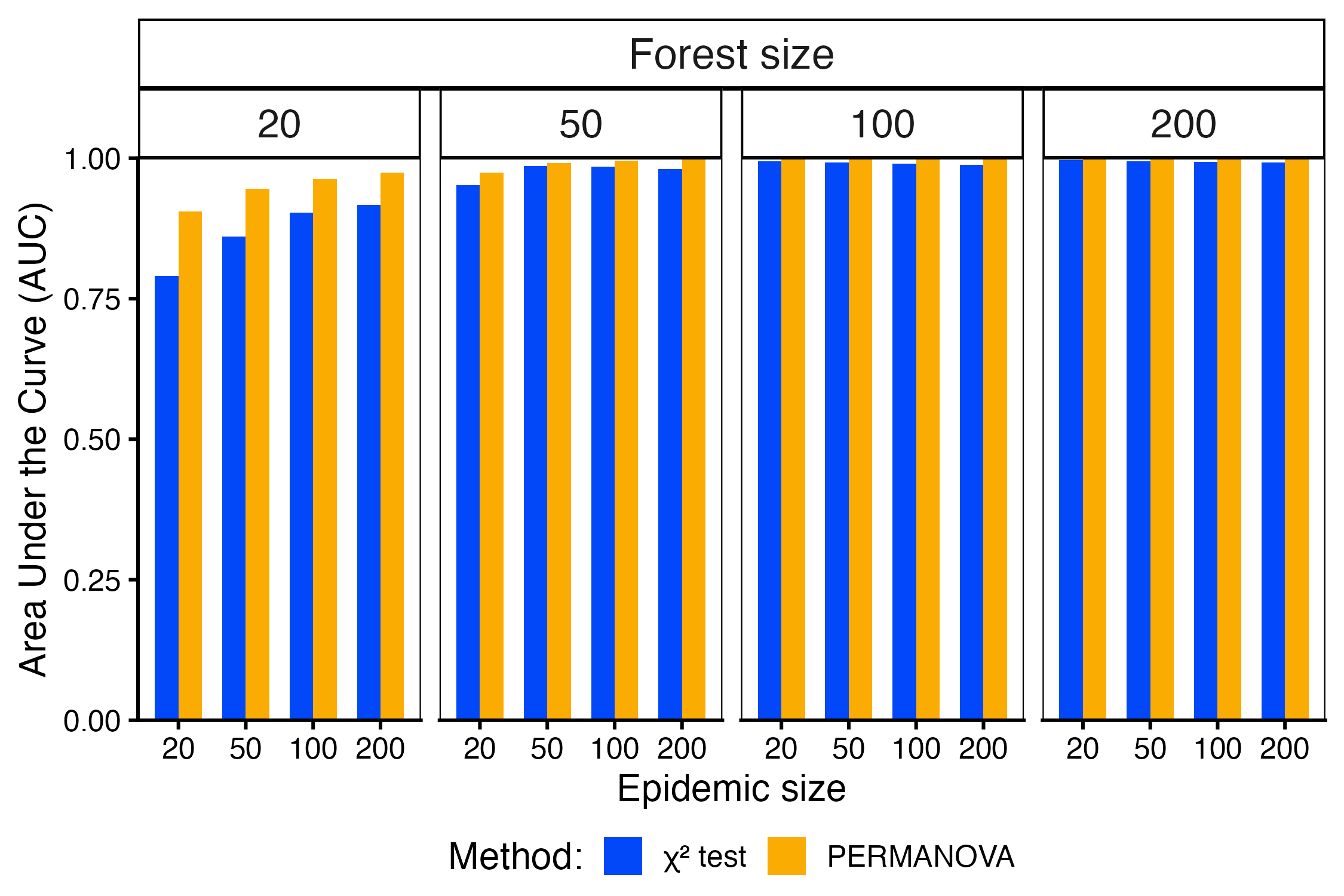}
    \caption{Area under curve (AUC) for $\chi^2$ test and PERMANOVA}
     \medskip
    \small
    \justifying
This figure shows the AUC derived from the Receiver Operating Characteristic (ROC) curves (supplementary Fig. \ref{fig:roc-curves}) of the two tests evaluated in our simulations. The y-axis displays the AUC value, with higher values corresponding to better performances. An AUC of 1 corresponds to a test with perfect sensitivity and specificity.  The x-axis displays the epidemic size \textit{i.e.} the number of cases in the simulated epidemics, while panels refer to the forest size \textit{i.e.} the number of trees in each forest.
    \label{fig:auc}
\end{figure}
\clearpage
\begin{table}[!h]
\centering
\begin{tabularx}{1\textwidth}{lrrrrrr} % or use adjustbox
\hline 
\textbf{method} & \textbf{min} & \makecell{\textbf{lower}\\\textbf{quartile}} & \textbf{mean} & \textbf{median} & \makecell{\textbf{upper}\\\textbf{quartile}} & \textbf{max} \\
\hline
PERMANOVA & 5.16 & 5.48 & 5.62 & 5.59 & 5.74 & 6.33 \\
$\chi^2$  test & 0.52 & 0.53 & 0.53 & 0.53 & 0.54 & 0.58 \\
\hline
\end{tabularx}
\caption{Benchmark results of execution times in seconds for the $\chi^2$ test and PERMANOVA, comparing two epidemic forests with 100 trees and 100 vertices each. Both tests used 999 permutations (PERMANOVA) / Monte Carlo replicates ($\chi^2$ test) without parallelisation and were replicated 100 times per method.}
\label{tab:benchmark_table}
\end{table}

For the $\chi^2$ test, we compute the frequency of each infector-infectee pair across all trees between forests. In the worst-case scenario, where every possible infector-infectee pair (\textit{i.e. }$n(n-1)$ pairs) appears at least once in either forest, the computational time for the $\chi^2$ test increases with the number of trees in each forest ($m$) and the square of the number of cases ($n^2$), since it considers all infector-infectee pairs for every tree (See Methods, Fig.\ref{fig:diagram}). Therefore the overall computational time will increase as a function of $mn^2$.

On the other hand, PERMANOVA involves a two-step process. First, it computes pairwise distances between all vertices within each tree (Fig. \ref{fig:patristic}), which scales as a function of $n^2$ for a given tree.  Second, it calculates pairwise distances between all trees (Fig.\ref{fig:diagram}), which scales as a function of $m^2$ . Therefore the overall computational time will increase as a function of $m^2n^2$.

%% file: references.bib
@misc{geismar_mixtree-analysis-data_2025,
	title = {mixtree-analysis-data},
	url = {https://zenodo.org/records/17704456},
	doi = {10.5281/zenodo.17704456},
	language = {eng},
	urldate = {2025-11-24},
	publisher = {Zenodo},
	author = {Geismar, Cyril},
	month = nov,
	year = {2025},
}

@misc{cygei_cygeimixtree_analysis_2025,
	title = {{CyGei}/mixtree\_analysis: mixtree-analysis},
	shorttitle = {{CyGei}/mixtree\_analysis},
	url = {https://zenodo.org/records/17704759},
	urldate = {2025-11-24},
	publisher = {Zenodo},
	author = {CyGei},
	month = nov,
	year = {2025},
	doi = {10.5281/zenodo.17704759},
}

@article{watson_probability_1875,
	title = {On the {Probability} of the {Extinction} of {Families}},
	volume = {4},
	issn = {09595295},
	url = {http://www.jstor.org/stable/2841222},
	doi = {10.2307/2841222},
	urldate = {2025-11-21},
	journal = {The Journal of the Anthropological Institute of Great Britain and Ireland},
	author = {Watson, H. W. and Galton, Francis},
	year = {1875},
	note = {Publisher: [Royal Anthropological Institute of Great Britain and Ireland, Wiley]},
	pages = {138--144},
}

@article{gudmundarson_gtst_2024,
	title = {{GTST}: {A} {Python} {Package} for {Graph} {Two}-{Sample} {Testing} {\textbar} {Journal} of {Open} {Research} {Software}},
	shorttitle = {{GTST}},
	url = {https://openresearchsoftware.metajnl.com/articles/10.5334/jors.478},
	doi = {10.5334/jors.478},
	language = {en},
	urldate = {2025-11-21},
	author = {Gudmundarson, Ragnar L. and Peters, Gareth W.},
	month = jan,
	year = {2024},
}

@article{shannon_mathematical_1948,
	title = {A mathematical theory of communication},
	volume = {27},
	issn = {0005-8580},
	number = {3},
	journal = {The Bell system technical journal},
	author = {Shannon, Claude E},
	year = {1948},
	note = {Publisher: Nokia Bell Labs},
	pages = {379--423},
}

@article{tjur_coefficients_2009,
	title = {Coefficients of {Determination} in {Logistic} {Regression} {Models}—{A} {New} {Proposal}: {The} {Coefficient} of {Discrimination}},
	volume = {63},
	issn = {0003-1305},
	shorttitle = {Coefficients of {Determination} in {Logistic} {Regression} {Models}—{A} {New} {Proposal}},
	url = {https://doi.org/10.1198/tast.2009.08210},
	doi = {10.1198/tast.2009.08210},
	number = {4},
	urldate = {2025-11-03},
	journal = {The American Statistician},
	author = {Tjur, Tue},
	month = nov,
	year = {2009},
	note = {Publisher: ASA Website
\_eprint: https://doi.org/10.1198/tast.2009.08210},
	pages = {366--372},
}

@article{jombart_treespace_2017,
	title = {treespace: {Statistical} exploration of landscapes of phylogenetic trees},
	volume = {17},
	copyright = {© 2017 The Authors. Molecular Ecology Resources Published by John Wiley \& Sons Ltd.},
	issn = {1755-0998},
	shorttitle = {treespace},
	url = {https://onlinelibrary.wiley.com/doi/abs/10.1111/1755-0998.12676},
	doi = {10.1111/1755-0998.12676},
	language = {en},
	number = {6},
	urldate = {2024-10-02},
	journal = {Molecular Ecology Resources},
	author = {Jombart, Thibaut and Kendall, Michelle and Almagro-Garcia, Jacob and Colijn, Caroline},
	year = {2017},
	note = {\_eprint: https://onlinelibrary.wiley.com/doi/pdf/10.1111/1755-0998.12676},
	pages = {1385--1392},
}

@article{bradley_monte_1977,
	title = {Monte {Carlo} simulations and the chi-square test of independence},
	volume = {9},
	issn = {1554-3528},
	url = {https://doi.org/10.3758/BF03214499},
	doi = {10.3758/BF03214499},
	language = {en},
	number = {2},
	urldate = {2025-03-24},
	journal = {Behavior Research Methods \& Instrumentation},
	author = {Bradley, Drake R. and Cutcomb, Steven},
	month = mar,
	year = {1977},
	pages = {193--201},
}

@article{r_core_team_r_2025,
	title = {R: {A} language and environment for statistical computing},
	author = {R Core Team, R},
	year = {2025},
	note = {Publisher: R foundation for statistical computing Vienna, Austria},
}

@misc{oksanen_vegan_2025,
	title = {vegan: {Community} {Ecology} {Package}},
	copyright = {GPL-2},
	shorttitle = {vegan},
	url = {https://cran.r-project.org/web/packages/vegan/index.html},
	urldate = {2025-03-20},
	author = {Oksanen, Jari and Simpson, Gavin L. and Blanchet, F. Guillaume and Kindt, Roeland and Legendre, Pierre and Minchin, Peter R. and O'Hara, R. B. and Solymos, Peter and Stevens, M. Henry H. and Szoecs, Eduard and Wagner, Helene and Barbour, Matt and Bedward, Michael and Bolker, Ben and Borcard, Daniel and Carvalho, Gustavo and Chirico, Michael and Caceres, Miquel De and Durand, Sebastien and Evangelista, Heloisa Beatriz Antoniazi and FitzJohn, Rich and Friendly, Michael and Furneaux, Brendan and Hannigan, Geoffrey and Hill, Mark O. and Lahti, Leo and McGlinn, Dan and Ouellette, Marie-Helene and Cunha, Eduardo Ribeiro and Smith, Tyler and Stier, Adrian and Braak, Cajo J. F. Ter and Weedon, James and Borman, Tuomas},
	month = jan,
	year = {2025},
}

@article{hall_epidemic_2015,
	title = {Epidemic {Reconstruction} in a {Phylogenetics} {Framework}: {Transmission} {Trees} as {Partitions} of the {Node} {Set}},
	volume = {11},
	issn = {1553-7358},
	shorttitle = {Epidemic {Reconstruction} in a {Phylogenetics} {Framework}},
	url = {https://journals.plos.org/ploscompbiol/article?id=10.1371/journal.pcbi.1004613},
	doi = {10.1371/journal.pcbi.1004613},
	language = {en},
	number = {12},
	urldate = {2025-03-15},
	journal = {PLOS Computational Biology},
	author = {Hall, Matthew and Woolhouse, Mark and Rambaut, Andrew},
	month = dec,
	year = {2015},
	note = {Publisher: Public Library of Science},
	pages = {e1004613},
}

@book{gibbons_algorithmic_1985,
	title = {Algorithmic {Graph} {Theory}},
	isbn = {978-0-521-28881-1},
	language = {en},
	publisher = {Cambridge University Press},
	author = {Gibbons, Alan},
	month = jun,
	year = {1985},
	note = {Google-Books-ID: Be6t04pgggwC},
}

@misc{geismar_mixtree_2025,
	title = {mixtree: {A} {Statistical} {Framework} for {Comparing} {Sets} of {Trees}},
	copyright = {MIT + file LICENSE},
	shorttitle = {mixtree},
	url = {https://cran.r-project.org/web/packages/mixtree/index.html},
	urldate = {2025-03-12},
	author = {Geismar, Cyril},
	month = mar,
	year = {2025},
}

@article{granovetter_strength_1973,
	title = {The {Strength} of {Weak} {Ties}},
	volume = {78},
	issn = {0002-9602},
	url = {https://www.jstor.org/stable/2776392},
	number = {6},
	urldate = {2025-03-12},
	journal = {American Journal of Sociology},
	author = {Granovetter, Mark S.},
	year = {1973},
	note = {Publisher: The University of Chicago Press},
	pages = {1360--1380},
}

@article{balaban_applications_1985,
	title = {Applications of graph theory in chemistry},
	volume = {25},
	issn = {0095-2338, 1520-5142},
	url = {https://pubs.acs.org/doi/abs/10.1021/ci00047a033},
	doi = {10.1021/ci00047a033},
	language = {en},
	number = {3},
	urldate = {2025-03-12},
	journal = {Journal of Chemical Information and Computer Sciences},
	author = {Balaban, Alexandru T.},
	month = aug,
	year = {1985},
	pages = {334--343},
}

@article{zerbino_velvet_2008,
	title = {Velvet: {Algorithms} for de novo short read assembly using de {Bruijn} graphs},
	volume = {18},
	issn = {1088-9051},
	shorttitle = {Velvet},
	url = {https://www.ncbi.nlm.nih.gov/pmc/articles/PMC2336801/},
	doi = {10.1101/gr.074492.107},
	number = {5},
	urldate = {2025-03-12},
	journal = {Genome Research},
	author = {Zerbino, Daniel R. and Birney, Ewan},
	month = may,
	year = {2008},
	pmid = {18349386},
	pmcid = {PMC2336801},
	pages = {821--829},
}

@article{chen_entity-relationship_1976,
	title = {The entity-relationship model—toward a unified view of data},
	volume = {1},
	issn = {0362-5915, 1557-4644},
	url = {https://dl.acm.org/doi/10.1145/320434.320440},
	doi = {10.1145/320434.320440},
	language = {en},
	number = {1},
	urldate = {2025-03-12},
	journal = {ACM Transactions on Database Systems},
	author = {Chen, Peter Pin-Shan},
	month = mar,
	year = {1976},
	pages = {9--36},
}

@book{gross_graph_2018,
	address = {New York},
	edition = {3},
	title = {Graph {Theory} and {Its} {Applications}},
	isbn = {978-0-429-42513-4},
	publisher = {Chapman and Hall/CRC},
	author = {Gross, Jonathan L. and Yellen, Jay and Anderson, Mark},
	month = nov,
	year = {2018},
	doi = {10.1201/9780429425134},
}

@article{felsenstein_confidence_1985,
	title = {{CONFIDENCE} {LIMITS} {ON} {PHYLOGENIES}: {AN} {APPROACH} {USING} {THE} {BOOTSTRAP}},
	volume = {39},
	issn = {0014-3820},
	shorttitle = {{CONFIDENCE} {LIMITS} {ON} {PHYLOGENIES}},
	url = {https://doi.org/10.1111/j.1558-5646.1985.tb00420.x},
	doi = {10.1111/j.1558-5646.1985.tb00420.x},
	number = {4},
	urldate = {2025-03-12},
	journal = {Evolution},
	author = {Felsenstein, Joseph},
	month = jul,
	year = {1985},
	pages = {783--791},
}

@article{kendall_estimating_2018,
	title = {Estimating {Transmission} from {Genetic} and {Epidemiological} {Data}: {A} {Metric} to {Compare} {Transmission} {Trees}},
	volume = {33},
	issn = {0883-4237},
	shorttitle = {Estimating {Transmission} from {Genetic} and {Epidemiological} {Data}},
	url = {https://www.jstor.org/stable/26770980},
	number = {1},
	urldate = {2024-10-02},
	journal = {Statistical Science},
	author = {Kendall, Michelle and Ayabina, Diepreye and Xu, Yuanwei and Stimson, James and Colijn, Caroline},
	year = {2018},
	note = {Publisher: Institute of Mathematical Statistics},
	pages = {70--85},
}

@incollection{bang-jensen_connectivity_2009,
	address = {London},
	title = {Connectivity of {Digraphs}},
	isbn = {978-1-84800-998-1},
	url = {https://doi.org/10.1007/978-1-84800-998-1_5},
	language = {en},
	urldate = {2025-03-09},
	booktitle = {Digraphs: {Theory}, {Algorithms} and {Applications}},
	publisher = {Springer},
	author = {Bang-Jensen, Jørgen and Gutin, Gregory Z.},
	editor = {Bang-Jensen, Jørgen and Gutin, Gregory Z.},
	year = {2009},
	doi = {10.1007/978-1-84800-998-1_5},
	pages = {191--226},
}

@article{lambert_students_2018,
	title = {A student's guide to {Bayesian} statistics},
	url = {https://www.torrossa.com/gs/resourceProxy?an=5017731&publisher=FZ7200},
	urldate = {2025-03-12},
	author = {Lambert, Ben},
	year = {2018},
	note = {Publisher: SAGE Publications Ltd},
}

@article{anderson_new_2001,
	title = {A new method for non-parametric multivariate analysis of variance},
	volume = {26},
	issn = {1442-9993},
	url = {https://onlinelibrary.wiley.com/doi/abs/10.1111/j.1442-9993.2001.01070.pp.x},
	doi = {10.1111/j.1442-9993.2001.01070.pp.x},
	language = {en},
	number = {1},
	urldate = {2025-03-09},
	journal = {Austral Ecology},
	author = {Anderson, Marti J.},
	year = {2001},
	note = {\_eprint: https://onlinelibrary.wiley.com/doi/pdf/10.1111/j.1442-9993.2001.01070.pp.x},
	pages = {32--46},
}

@article{colijn_metric_2018,
	title = {A {Metric} on {Phylogenetic} {Tree} {Shapes}},
	volume = {67},
	issn = {1063-5157},
	url = {https://www.ncbi.nlm.nih.gov/pmc/articles/PMC5790134/},
	doi = {10.1093/sysbio/syx046},
	number = {1},
	urldate = {2025-03-06},
	journal = {Systematic Biology},
	author = {Colijn, C. and Plazzotta, G.},
	month = jan,
	year = {2018},
	pmid = {28472435},
	pmcid = {PMC5790134},
	pages = {113--126},
}

@article{steel_distributions_1993,
	title = {Distributions of {Tree} {Comparison} {Metrics}—{Some} {New} {Results}},
	volume = {42},
	issn = {1063-5157},
	url = {https://doi.org/10.1093/sysbio/42.2.126},
	doi = {10.1093/sysbio/42.2.126},
	number = {2},
	urldate = {2025-03-06},
	journal = {Systematic Biology},
	author = {Steel, Mike A. and Penny, David},
	month = jun,
	year = {1993},
	pages = {126--141},
}

@inproceedings{robinson_comparison_1979,
	address = {Berlin, Heidelberg},
	title = {Comparison of weighted labelled trees},
	isbn = {978-3-540-34857-3},
	doi = {10.1007/BFb0102690},
	language = {en},
	booktitle = {Combinatorial {Mathematics} {VI}},
	publisher = {Springer},
	author = {Robinson, D. F. and Foulds, L. R.},
	editor = {Horadam, A. F. and Wallis, W. D.},
	year = {1979},
	pages = {119--126},
}

@article{pavoine_testing_2008,
	title = {Testing for phylogenetic signal in phenotypic traits: {New} matrices of phylogenetic proximities},
	volume = {73},
	issn = {0040-5809},
	shorttitle = {Testing for phylogenetic signal in phenotypic traits},
	url = {https://www.sciencedirect.com/science/article/pii/S0040580907001177},
	doi = {10.1016/j.tpb.2007.10.001},
	number = {1},
	urldate = {2025-03-06},
	journal = {Theoretical Population Biology},
	author = {Pavoine, Sandrine and Ollier, Sébastien and Pontier, Dominique and Chessel, Daniel},
	month = feb,
	year = {2008},
	pages = {79--91},
}

@article{drummond_beast_2007,
	title = {{BEAST}: {Bayesian} evolutionary analysis by sampling trees},
	volume = {7},
	issn = {1471-2148},
	shorttitle = {{BEAST}},
	url = {https://doi.org/10.1186/1471-2148-7-214},
	doi = {10.1186/1471-2148-7-214},
	number = {1},
	urldate = {2025-03-06},
	journal = {BMC Evolutionary Biology},
	author = {Drummond, Alexei J. and Rambaut, Andrew},
	month = nov,
	year = {2007},
	pages = {214},
}

@article{geismar_sorting_2024,
	title = {Sorting out assortativity: {When} can we assess the contributions of different population groups to epidemic transmission?},
	volume = {19},
	issn = {1932-6203},
	shorttitle = {Sorting out assortativity},
	url = {https://journals.plos.org/plosone/article?id=10.1371/journal.pone.0313037},
	doi = {10.1371/journal.pone.0313037},
	language = {en},
	number = {12},
	urldate = {2024-12-02},
	journal = {PLOS ONE},
	author = {Geismar, Cyril and White, Peter J. and Cori, Anne and Jombart, Thibaut},
	month = dec,
	year = {2024},
	note = {Publisher: Public Library of Science},
	pages = {e0313037},
}

@article{lloyd-smith_superspreading_2005,
	title = {Superspreading and the effect of individual variation on disease emergence},
	volume = {438},
	copyright = {2005 Springer Nature Limited},
	issn = {1476-4687},
	url = {https://www.nature.com/articles/nature04153},
	doi = {10.1038/nature04153},
	number = {7066},
	urldate = {2024-09-02},
	journal = {Nature},
	author = {Lloyd-Smith, J. O. and Schreiber, S. J. and Kopp, P. E. and Getz, W. M.},
	month = nov,
	year = {2005},
	note = {Publisher: Nature Publishing Group},
	pages = {355--359},
}

@article{frieden_identifying_2020,
	title = {Identifying and {Interrupting} {Superspreading} {Events}—{Implications} for {Control} of {Severe} {Acute} {Respiratory} {Syndrome} {Coronavirus} 2},
	volume = {26},
	url = {https://pmc.ncbi.nlm.nih.gov/articles/PMC7258476/},
	doi = {10.3201/eid2606.200495},
	language = {en},
	number = {6},
	urldate = {2024-10-23},
	journal = {Emerging Infectious Diseases},
	author = {Frieden, Thomas R. and Lee, Christopher T.},
	month = jun,
	year = {2020},
	pmid = {32187007},
	pages = {1059},
}

@article{pearson_x_1900,
	title = {X. {On} the criterion that a given system of deviations from the probable in the case of a correlated system of variables is such that it can be reasonably supposed to have arisen from random sampling},
	volume = {50},
	issn = {1941-5982},
	number = {302},
	journal = {The London, Edinburgh, and Dublin Philosophical Magazine and Journal of Science},
	author = {Pearson, Karl},
	year = {1900},
	note = {Publisher: Taylor \& Francis},
	pages = {157--175},
}

@article{rai_estimates_2021,
	title = {Estimates of serial interval for {COVID}-19: {A} systematic review and meta-analysis},
	volume = {9},
	issn = {2213-3984},
	shorttitle = {Estimates of serial interval for {COVID}-19},
	url = {https://www.sciencedirect.com/science/article/pii/S2213398420301895},
	doi = {10.1016/j.cegh.2020.08.007},
	urldate = {2024-10-02},
	journal = {Clinical Epidemiology and Global Health},
	author = {Rai, Balram and Shukla, Anandi and Dwivedi, Laxmi Kant},
	month = jan,
	year = {2021},
	pages = {157--161},
}

@article{kremer_reconstruction_2023,
	title = {Reconstruction of {SARS}-{CoV}-2 outbreaks in a primary school using epidemiological and genomic data},
	volume = {44},
	issn = {1755-4365},
	url = {https://www.ncbi.nlm.nih.gov/pmc/articles/PMC10273772/},
	doi = {10.1016/j.epidem.2023.100701},
	urldate = {2024-09-10},
	journal = {Epidemics},
	author = {Kremer, Cécile and Torneri, Andrea and Libin, Pieter J.K. and Meex, Cécile and Hayette, Marie-Pierre and Bontems, Sébastien and Durkin, Keith and Artesi, Maria and Bours, Vincent and Lemey, Philippe and Darcis, Gilles and Hens, Niel and Meuris, Christelle},
	month = sep,
	year = {2023},
	pmid = {37379776},
	pmcid = {PMC10273772},
	pages = {100701},
}

@article{wang_inference_2020,
	title = {Inference of person-to-person transmission of {COVID}-19 reveals hidden super-spreading events during the early outbreak phase},
	volume = {11},
	issn = {2041-1723},
	number = {1},
	journal = {Nature communications},
	author = {Wang, Liang and Didelot, Xavier and Yang, Jing and Wong, Gary and Shi, Yi and Liu, Wenjun and Gao, George F and Bi, Yuhai},
	year = {2020},
	note = {Publisher: Nature Publishing Group UK London},
	pages = {5006},
}

@article{de_maio_scotti_2016,
	title = {{SCOTTI}: efficient reconstruction of transmission within outbreaks with the structured coalescent},
	volume = {12},
	issn = {1553-734X},
	number = {9},
	journal = {PLoS computational biology},
	author = {De Maio, Nicola and Wu, Chieh-Hsi and Wilson, Daniel J},
	year = {2016},
	note = {Publisher: Public Library of Science San Francisco, CA USA},
	pages = {e1005130},
}

@article{klinkenberg_simultaneous_2017,
	title = {Simultaneous inference of phylogenetic and transmission trees in infectious disease outbreaks},
	volume = {13},
	issn = {1553-734X},
	number = {5},
	journal = {PLoS computational biology},
	author = {Klinkenberg, Don and Backer, Jantien A and Didelot, Xavier and Colijn, Caroline and Wallinga, Jacco},
	year = {2017},
	note = {Publisher: Public Library of Science San Francisco, CA USA},
	pages = {e1005495},
}

@article{didelot_bayesian_2014,
	title = {Bayesian {Inference} of {Infectious} {Disease} {Transmission} from {Whole}-{Genome} {Sequence} {Data}},
	volume = {31},
	issn = {0737-4038},
	url = {https://doi.org/10.1093/molbev/msu121},
	doi = {10.1093/molbev/msu121},
	number = {7},
	urldate = {2024-09-05},
	journal = {Molecular Biology and Evolution},
	author = {Didelot, Xavier and Gardy, Jennifer and Colijn, Caroline},
	month = jul,
	year = {2014},
	pages = {1869--1879},
}

@article{abbas_reconstruction_2022,
	title = {Reconstruction of transmission chains of {SARS}-{CoV}-2 amidst multiple outbreaks in a geriatric acute-care hospital: a combined retrospective epidemiological and genomic study},
	volume = {11},
	issn = {2050-084X},
	shorttitle = {Reconstruction of transmission chains of {SARS}-{CoV}-2 amidst multiple outbreaks in a geriatric acute-care hospital},
	url = {https://doi.org/10.7554/eLife.76854},
	doi = {10.7554/eLife.76854},
	urldate = {2024-02-23},
	journal = {eLife},
	author = {Abbas, Mohamed and Cori, Anne and Cordey, Samuel and Laubscher, Florian and Robalo Nunes, Tomás and Myall, Ashleigh and Salamun, Julien and Huber, Philippe and Zekry, Dina and Prendki, Virginie and Iten, Anne and Vieux, Laure and Sauvan, Valérie and Graf, Christophe E and Harbarth, Stephan},
	editor = {Schiffer, Joshua T and Garrett, Wendy S and Opatowski, Lulla and Pham, Thi},
	month = jul,
	year = {2022},
	note = {Publisher: eLife Sciences Publications, Ltd},
	pages = {e76854},
}

@article{duault_methods_2022,
	title = {Methods {Combining} {Genomic} and {Epidemiological} {Data} in the {Reconstruction} of {Transmission} {Trees}: {A} {Systematic} {Review}},
	volume = {11},
	shorttitle = {Methods {Combining} {Genomic} and {Epidemiological} {Data} in the {Reconstruction} of {Transmission} {Trees}},
	url = {https://www.mdpi.com/2076-0817/11/2/252},
	number = {2},
	urldate = {2024-01-04},
	journal = {Pathogens},
	author = {Duault, Hélène and Durand, Benoit and Canini, Laetitia},
	year = {2022},
	note = {Publisher: MDPI},
	pages = {252},
}

@article{cori_new_2013,
	title = {A {New} {Framework} and {Software} to {Estimate} {Time}-{Varying} {Reproduction} {Numbers} {During} {Epidemics}},
	volume = {178},
	issn = {0002-9262},
	url = {https://doi.org/10.1093/aje/kwt133},
	doi = {10.1093/aje/kwt133},
	number = {9},
	urldate = {2024-08-08},
	journal = {American Journal of Epidemiology},
	author = {Cori, Anne and Ferguson, Neil M. and Fraser, Christophe and Cauchemez, Simon},
	month = nov,
	year = {2013},
	pages = {1505--1512},
}

@article{vink_serial_2014,
	title = {Serial {Intervals} of {Respiratory} {Infectious} {Diseases}: {A} {Systematic} {Review} and {Analysis}},
	volume = {180},
	issn = {0002-9262},
	shorttitle = {Serial {Intervals} of {Respiratory} {Infectious} {Diseases}},
	url = {https://doi.org/10.1093/aje/kwu209},
	doi = {10.1093/aje/kwu209},
	number = {9},
	urldate = {2024-08-08},
	journal = {American Journal of Epidemiology},
	author = {Vink, Margaretha Annelie and Bootsma, Martinus Christoffel Jozef and Wallinga, Jacco},
	month = nov,
	year = {2014},
	pages = {865--875},
}

@article{gelman_inference_1992,
	title = {Inference from {Iterative} {Simulation} {Using} {Multiple} {Sequences}},
	volume = {7},
	issn = {0883-4237, 2168-8745},
	url = {https://projecteuclid.org/journals/statistical-science/volume-7/issue-4/Inference-from-Iterative-Simulation-Using-Multiple-Sequences/10.1214/ss/1177011136.full},
	doi = {10.1214/ss/1177011136},
	number = {4},
	urldate = {2024-07-18},
	journal = {Statistical Science},
	author = {Gelman, Andrew and Rubin, Donald B.},
	month = nov,
	year = {1992},
	note = {Publisher: Institute of Mathematical Statistics},
	pages = {457--472},
}

@article{jombart_reconstructing_2011,
	title = {Reconstructing disease outbreaks from genetic data: a graph approach},
	volume = {106},
	copyright = {2011 The Genetics Society},
	issn = {1365-2540},
	shorttitle = {Reconstructing disease outbreaks from genetic data},
	url = {https://www.nature.com/articles/hdy201078},
	doi = {10.1038/hdy.2010.78},
	language = {en},
	number = {2},
	urldate = {2024-07-18},
	journal = {Heredity},
	author = {Jombart, T. and Eggo, R. M. and Dodd, P. J. and Balloux, F.},
	month = feb,
	year = {2011},
	note = {Publisher: Nature Publishing Group},
	pages = {383--390},
}

@article{campbell_bayesian_2019,
	title = {Bayesian inference of transmission chains using timing of symptoms, pathogen genomes and contact data},
	volume = {15},
	issn = {1553-7358},
	url = {https://journals.plos.org/ploscompbiol/article?id=10.1371/journal.pcbi.1006930},
	doi = {10.1371/journal.pcbi.1006930},
	language = {en},
	number = {3},
	urldate = {2024-07-18},
	journal = {PLOS Computational Biology},
	author = {Campbell, Finlay and Cori, Anne and Ferguson, Neil and Jombart, Thibaut},
	month = mar,
	year = {2019},
	note = {Publisher: Public Library of Science},
	pages = {e1006930},
}

@article{didelot_genomic_2017,
	title = {Genomic {Infectious} {Disease} {Epidemiology} in {Partially} {Sampled} and {Ongoing} {Outbreaks}},
	volume = {34},
	issn = {0737-4038},
	url = {https://doi.org/10.1093/molbev/msw275},
	doi = {10.1093/molbev/msw275},
	number = {4},
	urldate = {2024-07-13},
	journal = {Molecular Biology and Evolution},
	author = {Didelot, Xavier and Fraser, Christophe and Gardy, Jennifer and Colijn, Caroline},
	month = apr,
	year = {2017},
	pages = {997--1007},
}

@article{campbell_when_2018,
	title = {When are pathogen genome sequences informative of transmission events?},
	volume = {14},
	issn = {1553-7374},
	url = {https://journals.plos.org/plospathogens/article?id=10.1371/journal.ppat.1006885},
	doi = {10.1371/journal.ppat.1006885},
	language = {en},
	number = {2},
	urldate = {2024-06-13},
	journal = {PLOS Pathogens},
	author = {Campbell, Finlay and Strang, Camilla and Ferguson, Neil and Cori, Anne and Jombart, Thibaut},
	month = feb,
	year = {2018},
	note = {Publisher: Public Library of Science},
	pages = {e1006885},
}

@article{jombart_bayesian_2014,
	title = {Bayesian {Reconstruction} of {Disease} {Outbreaks} by {Combining} {Epidemiologic} and {Genomic} {Data}},
	volume = {10},
	issn = {1553-7358},
	url = {https://journals.plos.org/ploscompbiol/article?id=10.1371/journal.pcbi.1003457},
	doi = {10.1371/journal.pcbi.1003457},
	language = {en},
	number = {1},
	urldate = {2024-01-30},
	journal = {PLOS Computational Biology},
	author = {Jombart, Thibaut and Cori, Anne and Didelot, Xavier and Cauchemez, Simon and Fraser, Christophe and Ferguson, Neil},
	month = jan,
	year = {2014},
	note = {Publisher: Public Library of Science},
	pages = {e1003457},
}

@article{campbell_outbreaker2_2018,
	title = {outbreaker2: a modular platform for outbreak reconstruction},
	volume = {19},
	issn = {1471-2105},
	shorttitle = {outbreaker2},
	url = {https://bmcbioinformatics.biomedcentral.com/articles/10.1186/s12859-018-2330-z},
	doi = {10.1186/s12859-018-2330-z},
	language = {en},
	number = {S11},
	urldate = {2024-01-05},
	journal = {BMC Bioinformatics},
	author = {Campbell, Finlay and Didelot, Xavier and Fitzjohn, Rich and Ferguson, Neil and Cori, Anne and Jombart, Thibaut},
	month = oct,
	year = {2018},
	pages = {363},
}

@article{geismar_bayesian_2023,
	title = {Bayesian reconstruction of {SARS}-{CoV}-2 transmissions highlights substantial proportion of negative serial intervals},
	volume = {44},
	url = {https://www.sciencedirect.com/science/article/pii/S175543652300049X},
	urldate = {2024-01-05},
	journal = {Epidemics},
	author = {Geismar, Cyril and Nguyen, Vincent and Fragaszy, Ellen and Shrotri, Madhumita and Navaratnam, Annalan MD and Beale, Sarah and Byrne, Thomas E. and Fong, Wing Lam Erica and Yavlinsky, Alexei and Kovar, Jana},
	year = {2023},
	note = {Publisher: Elsevier},
	pages = {100713},
}

@article{abbas_explosive_2021,
	title = {Explosive nosocomial outbreak of {SARS}-{CoV}-2 in a rehabilitation clinic: the limits of genomics for outbreak reconstruction},
	volume = {117},
	shorttitle = {Explosive nosocomial outbreak of {SARS}-{CoV}-2 in a rehabilitation clinic},
	url = {https://www.sciencedirect.com/science/article/pii/S019567012100308X},
	urldate = {2024-01-04},
	journal = {Journal of Hospital Infection},
	author = {Abbas, Mohamed and Nunes, T. Robalo and Cori, Anne and Cordey, Samuel and Laubscher, Florian and Baggio, Stephanie and Jombart, Thibaut and Iten, Anne and Vieux, Laure and Teixeira, Daniel},
	year = {2021},
	note = {Publisher: Elsevier},
	pages = {124--134},
}
